\documentclass[12pt,epsf,qsymbols]{article}
\pdfoutput=1
\usepackage[utf8]{inputenc}
\usepackage[normalem]{ulem}
\usepackage[english]{babel}%,frenchb
\usepackage{tabularx}
\usepackage{array}
\usepackage{graphics}
\usepackage[pdftex]{graphicx}
\usepackage{psfrag}
\usepackage{epsfig}
\usepackage{subfig}
\usepackage{mathtools}
\usepackage{amssymb}
\usepackage{pifont}
\usepackage{bbold}
\usepackage{setspace}
\usepackage{rotating}
\usepackage{colortbl}
\usepackage{longtable}
\usepackage{slashed}
\usepackage{braket}
\usepackage{lineno}
\makeatletter
\usepackage{textcomp}
\usepackage[usenames,dvipsnames]{xcolor}
\usepackage{relsize}
\usepackage{cite}

\usepackage{float}
\usepackage{multirow}
\usepackage{tikz}
\usetikzlibrary{decorations.pathmorphing,shapes}

\usepackage{lipsum}
\usepackage{tabu}

\usepackage{bbm}
\usepackage{bm}
\usepackage{enumitem}
\usepackage{amsmath}
\usepackage{verbatim}

\usepackage{hyperref}
\hypersetup{colorlinks,citecolor=nicegreen,linkcolor=niceblue}
\hypersetup{colorlinks=true}

\usepackage{pifont}% http://ctan.org/pkg/pifont

\usepackage{calligra}

\usepackage{geometry}

\allowdisplaybreaks

\setlongtables

\newcommand{\bea}{\begin{eqnarray}}
\newcommand{\eea}{\end{eqnarray}}
\newcommand{\beq}{\begin{equation}}

\newcommand{\eeq}{\end{equation}}
\newcommand{\ec}{\end{center}}
\newcommand{\bc}{\begin{center}}

\newcommand{\pdir}{p\kern -5.2pt\raise 0.2ex\hbox {/}}

\newcommand{\vdir}{v\kern -5.75pt\raise 0.15ex\hbox {/}}
\newcommand{\kdir}{k\kern -5.75pt\raise 0.15ex\hbox {/}}
\newcommand{\epsdir}{\epsilon\kern -5.0pt\raise 0.15ex\hbox {/}}
\newcommand{\bvdir}{\bar{v}\kern -5.75pt\raise 0.15ex\hbox {/}}
\newcommand{\Ddir}{D\kern -7.75pt\raise 0.20ex\hbox {/}}
\newcommand{\Adir}{A\kern -7.75pt\raise 0.20ex\hbox {/}}
\newcommand{\ldir}{l\kern -5.0pt\raise 0.2ex\hbox{/}}
\newcommand{\varepsdir}{\varepsilon\kern -5.5pt\raise 0.15ex\hbox{/}}

\makeatother

\DeclareMathAlphabet{\mathcalligra}{T1}{calligra}{m}{n}
\DeclareFontShape{T1}{calligra}{m}{n}{<->s*[2.2]callig15}{}

\definecolor{lightgray}{rgb}{0.9,0.9,0.9}
\definecolor{niceblue}{rgb}{0.15,0.15,0.6}
\definecolor{nicegreen}{rgb}{0.1,0.5,0.1}
\definecolor{Red}{rgb}{1.,0.,0.}

\definecolor{Green}{rgb}{0.2,.7,0.2}

\begin{document}
\unitlength = 1mm

\thispagestyle{empty} 
\begin{center}
\vskip 1.8cm\par
{\par\centering \textbf{\LARGE  
\Large \bf Correlating $0\nu\beta\beta$ decays and flavor observables in \\[0.3em] leptoquark models}}

\vskip 1.2cm\par
{\scalebox{.85}{\par\centering \large  
\sc S.~Fajfer$^{\,a}$, L.P.S.~Leal$^{\,b,c}$, O.~Sumensari$^{\,b}$, R.~Zukanovich Funchal$^{\,c}$ }
{\par\centering \vskip 0.7 cm\par}
{\sl 
$^a$~
{\small Jožef Stefan Institute, Jamova 39, P. O. Box 3000, 1001 Ljubljana, Slovenia}\\[0.3em]
$^b$~
{\small Universit\'e Paris-Saclay, CNRS/IN2P3, IJCLab, 91405 Orsay, France}}\\[0.3em]
$^c$~
{\small Departamento de Física Matemática, Instituto de Física,\\ Universidade de S\~{a}o Paulo, 05315-970 S\~{a}o Paulo, Brazil}
}

{\vskip 1.65cm\par}
\end{center}

\vskip 0.85cm

\begin{abstract}
In this paper, we investigate minimal scalar leptoquark models that dynamically generate neutrino Majorana masses at the one-loop level and examine their implications for low-energy processes. We show that these models can produce viable neutrino masses, consistent with neutrino oscillation and cosmological data. 
By using leptoquark couplings fixed by neutrino data, we predict additional contributions to neutrinoless double-beta decays ($0\nu\beta\beta$), which are chirality enhanced and compete with the standard contributions from the Majorana masses. Our analysis demonstrates that these effects are sizable for leptoquark masses as large as $\mathcal{O}(300~\mathrm{TeV})$, potentially increasing or decreasing the $0\nu\beta\beta$ half-life, and creating an ambiguity between the normal and inverted mass ordering scenarios. Furthermore, we explore the correlation between $0\nu\beta\beta$ and flavor observables, such as kaon decays and $\mu\to e$ conversion in nuclei, emphasizing that the latter is complementary to $0\nu\beta\beta$ decays.

\end{abstract}

%\newpage
\setcounter{page}{1}
\setcounter{footnote}{0}
\setcounter{equation}{0}
%%%%%%%%%%%%%%%%%%%%%%%%%%%%%%%%%%%%%%%%
\noindent

\renewcommand{\thefootnote}{\arabic{footnote}}
%\linenumbers
 
\setcounter{footnote}{0}

%\clearpage

%\tableofcontents

\newpage

%%%%%%%%%%%%%%%%%%%%%%%%%%%%%%%%%%%%%%%%
%%%%%%%%%%%%%%%%%%%%%%%%%%%%%%%%%%%%%%%%
\section{Introduction}\label{sec:intro}
%%%%%%%%%%%%%%%%%%%%%%%%%%%%%%%%%%%%%%%%
%%%%%%%%%%%%%%%%%%%%%%%%%%%%%%%%%%%%%%%%

The observation of neutrino oscillation is unambiguous evidence of physics beyond the Standard Model (SM). Experiments dedicated to study solar, atmospheric, accelerator, and reactor neutrinos have established that neutrinos have tiny masses and that they oscillate among different flavors~\cite{SajjadAthar:2021prg}. However, the origin of such tiny masses remains unknown and several questions regarding the neutrino sector remain unanswered. These include the ambiguity between normal and inverted ordering for neutrino masses, and the determination if CP is violated in the lepton sector, which still cannot be resolved with current oscillation data, and the even more fundamental question if neutrinos are Dirac or Majorana particles.

The existence of Majorana neutrinos is theoretically compelling since the small neutrino masses ($m_{\nu_i}$) can be described by the only dimension-five operator allowed by the SM gauge symmetry~\cite{Weinberg:1979sa}. The suppression of this operator by the inverse power of the Effective Field Theory (EFT) cutoff $\Lambda$ would then explain the smallness of $m_{\nu_i}$. Since this operator violates Lepton Number ($L$), the observation of processes that violate $L$ by two units such as neutrinoless double-beta decays ($0\nu\beta\beta$) would be a clean indication of the Majorana neutrino nature. Several experiments have set strong upper-limits on the $0\nu\beta\beta$ half-life over the past years~\cite{KamLAND-Zen:2024eml,GERDA:2020xhi}, which will be further improved by the next-generation experiments, with an expected sensitivity of $10^{27}$--$10^{28}$ years~\cite{LEGEND:2021bnm,nEXO:2021ujk}. Given the theoretical estimations of the needed nuclear matrix-elements~\cite{Menendez:2008jp,Hyvarinen:2015bda,Barea:2015kwa,Menendez:2017fdf,Deppisch:2020ztt}, these future experiments will have the potential to observe $0\nu\beta\beta$ decays if neutrino masses are induced by the standard $d=5$ Weinberg operator and, in particular,  if they follow the inverted-mass ordering~\cite{Dolinski:2019nrj}.

The interpretation of experimental results on the $0\nu\beta\beta$ half-life can be more involved in scenarios where contributions beyond the Weinberg operator are present. This is the case in dynamical scenarios that generate neutrino masses at loop level, as they typically generate other higher-dimensional operators that also violate $L$~\cite{Cirigliano:2022oqy,Gargalionis:2020xvt}. The latter contributions can be relevant if the additional $(v/\Lambda)^n$ suppression is compensated by chirality-enhanced contributions $(\propto E/m_{\nu_i})$, in comparison to the standard ones, where $E\approx 100~\mathrm{MeV}$ is the typical nuclear energy-scale of these low-energy processes~\cite{Prezeau:2003xn}. A prime example of such models are the ones involving TeV-scale scalar leptoquarks~\cite{Chua:1999si}, which induce loop-level neutrino masses and tree-level contributions to $0\nu\beta\beta$, which are chirality enhanced~\cite{Hirsch:1996qy} (see also Ref.~\cite{Graf:2022lhj,Scholer:2023bnn}). For such scenarios, the interpretation of the experimental measurements in terms of the effective Majorana mass ($m_{\beta\beta}$) is more involved, and the conclusion that the next-generation experiments are capable of either excluding or observing the inverted-ordering scenario might no longer be valid.

In this paper, we consider the simplest leptoquark models that can generate neutrino masses at one-loop level~\cite{AristizabalSierra:2007nf}, and we perform a detailed study of the correlation between the loop-induced neutrino masses and the $0\nu\beta\beta$ half-life, which has not been systematically studied before. We will demonstrate that the predictions for $0\nu\beta\beta$ decays can deviate from the standard ones for leptoquark masses below $\approx 300$~TeV, creating an ambiguity if one aims to infer the neutrino mass ordering by only using the information from the $0\nu\beta\beta$ half-life.~\footnote{Similar conclusions have been obtained, e.g.,~in Ref.~\cite{Abada:2018qok} by extending the SM with light sterile-neutrinos, and in Ref.~\cite{Tello:2010am} in left-right symmetric models.} Furthermore, we will show that flavor-physics observables in the kaon sector, such as $K\to \pi\nu\nu$, and forbidden processes in the SM, such as $K_L\to \mu e$ and $\mu\to e$ conversion in nuclei, can already probe a comparable mass range, providing complementary constraints to $0\nu\beta\beta$ decays. In particular, we will show that future experiments searching for $\mu N\to e N$ will cover a substantial fraction of the leptoquark parameter space that will be accessible in the next generation of $0\nu\beta\beta$ experiments.~\footnote{See Ref.~\cite{Cirigliano:2004tc} where a similar conclusion was reached in different scenarios for neutrino masses.}

The remainder of this paper is organized as follows. In Sec.~\ref{sec:lq-model}, we introduce the two minimal models with scalar leptoquarks that generate neutrino masses at one loop. In Sec.~\ref{sec:0nubb}, we introduce the low-energy EFT describing $0\nu\beta\beta$ decays, which is matched onto the 
SMEFT and the leptoquark models. In Sec.~\ref{sec:pheno}, we discuss the main theoretical and phenomenological constraints on these scenarios, and we present our numerical results in Sec.~\ref{sec:numerical}. Finally, our findings are summarized in Sec.~\ref{sec:conclusion}.

%%%%%%%%%%%%%%%%%%%%%%%%%%%%%%%%%%%%%%%%
%%%%%%%%%%%%%%%%%%%%%%%%%%%%%%%%%%%%%%%%
\section{Leptoquark mechanism for $m_\nu$}\label{sec:lq-model}
%%%%%%%%%%%%%%%%%%%%%%%%%%%%%%%%%%%%%%%%
%%%%%%%%%%%%%%%%%%%%%%%%%%%%%%%%%%%%%%%%

Lepton Number Violation (LNV) can arise through the mixing of a pair of scalar leptoquarks carrying different values for the fermion number $F=3B+L$ (i.e.,~$F=0$ or $2$)~\cite{Buchmuller:1986zs}, where $B$ and $L$ denote baryon and lepton number, respectively. Under the assumption that the scalar sector only comprises the SM Higgs in addition to the leptoquarks, there are only two possibilities to break $L$ in the scalar sector: the weak doublet $\widetilde{R}_2 \sim (\mathbf{3},\mathbf{2},1/6)$ has to be combined with either (i) ${S}_1 \sim (\mathbf{\bar{3}},\mathbf{1},1/3)$, or (ii) ${S}_3 \sim (\mathbf{\bar{3}},\mathbf{3},1/3)$~\cite{Chua:1999si}.~\footnote{The mixing of two leptoquarks with same values of $F$ conserves $L$ and does not contribute to neutrino masses, but does contribute to other types of chirality-flipping observables such as leptonic dipoles~\cite{Dorsner:2019itg}.} These states are specified by their SM quantum numbers $(SU(3)_c,SU(2)_L,U(1)_Y)$, where the electric charge, $Q = Y + T_3$, is the sum of the hypercharge ($Y$) and the third-component of the weak isospin ($T_3$).  

\subsection{Scalar leptoquark models}

In the following, we describe the two minimal scalar leptoquark models that can generate neutrino masses. We will not write the quartic operators involving leptoquarks, since they do not play an important role in the following discussion (see e.g.~Ref.\cite{Crivellin:2021ejk} for a complete list of operators). Furthermore, we assume that the diquark couplings of $S_1$ and $S_3$ are forbidden by a suitable symmetry to ensure the proton stability. Notice that $\widetilde{R}_2$ automatically preserves baryon number at the renormalizable level thanks to the SM gauge symmetry~\cite{Assad:2017iib}.

\subsubsection{$\widetilde{R}_2$--$S_1$}
The Lagrangian of the first model reads
%%%%%%%%%%%%%%
\begin{align}
\label{eq:lag-R2t-S1}
\mathcal{L}_{\widetilde{R}_2 \& S_1} &\supset (D_\mu \widetilde{R}_2)^\dagger (D_\mu \widetilde{R}_2) + (D_\mu {S}_1)^\dagger (D_\mu {S}_1) \\[0.35em]
&-m_{\widetilde{R}_2}^2\, \widetilde{R}_2^\dagger \widetilde{R}_2 -m_{{S}_1}^2\, {S}_1^\dagger {S}_1 +\mathcal{L}_\mathrm{Yuk.} - \big{\lbrace}\lambda_{1}\, \widetilde{R}_2^\dagger H S_1^\ast +\mathrm{h.c.}\big{\rbrace} \,,\nonumber
\end{align}
%%%%%%%%%%%%%%

\noindent where $\lambda_{1}$ is the trilinear coupling that mixes the leptoquark states and that will be responsible for LNV. The Yukawa interactions are given by
%%%%%%%%%%%%%%
\begin{align}
\begin{split}
\label{eq:R2t-Yuk}
\mathcal{L}^{\widetilde{R}_2}_\mathrm{Yuk.} &=-y_{2L}^{ij}\,\bar{d}_{Ri} \widetilde{R}_2 i\tau_2 L_j+\mathrm{h.c.}\,,\\[0.4em]
\mathcal{L}^{S_1}_\mathrm{Yuk.}&=y_{1L}^{ij}\, \overline{Q^C_i} i\tau_2 L_j \,S_1 +y_{1R}^{ij}\, \overline{u^C_{Ri}} \ell_{Rj} \,S_1+\mathrm{h.c.}\,,
\end{split}
\end{align}
%%%%%%%%%%%%%%

%%%%%%%%%%%%%%
\begin{figure}[!t]
\begin{center}
\includegraphics[width=0.33\textwidth]{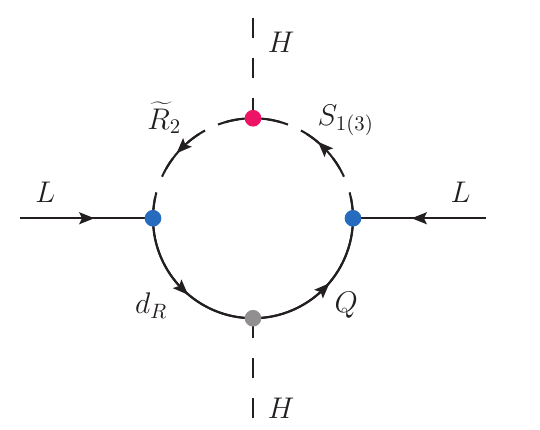} 
\end{center}
\caption{\sl\small Leptoquark contributions to the $d=5$ Weinberg operator~\cite{Weinberg:1979sa} via the mixing of $\widetilde{R}_2\sim (\mathbf{3},\mathbf{2},1/6)$ with $S_1\sim (\mathbf{\bar{3}},\mathbf{1},1/3)$ or $S_3\sim (\mathbf{\bar{3}},\mathbf{3},1/3)$.
\label{fig:diag-loop}
}
\end{figure}
%%%%%%%%%%%%%%

\noindent where  $i,j$ denote flavor indices, $\psi^C$ stand for the conjugate of the fermion field $\psi$, and the SM lepton and quark doublets are denoted by $Q$ and $L$, respectively, whereas the weak singlets are denoted by $\ell_R$, $u_R$ and $d_R$. The leptoquark Yukawa couplings are denoted by $\smash{y_{2L}}$, $y_{1L}$ and $y_{1R}$, which are written in the interaction basis, cf.~Sec.~\ref{sec:leptonic-mixing}.

\subsubsection{$\widetilde{R}_2$--$S_3$}
 
Similarly, the Lagrangian for the second model reads
%%%%%%%%%%%%%%
\begin{align}
\label{eq:lag-R2t-S3}
\mathcal{L}_{\widetilde{R}_2 \& S_3} &\supset (D_\mu \widetilde{R}_2)^\dagger (D_\mu \widetilde{R}_2) + (D_\mu {S}_3^a)^\dagger (D_\mu {S}_3^a) \\[0.4em]
&-m_{\widetilde{R}_2}^2\, \widetilde{R}_2^\dagger \widetilde{R}_2 -m_{{S}_3}^2\, {S}_3^{a\,\dagger} {S}_3^a +\mathcal{L}_\mathrm{Yuk.} - \big{\lbrace}\lambda_{3}\, \widetilde{R}_2^\dagger ({\tau}^a\,S_3^{a})^\dagger H +\mathrm{h.c.}\big{\rbrace} \, ,\nonumber
\end{align}
%%%%%%%%%%%%%%

\noindent where $\tau^a$ ($a=1,2,3$) denote the Pauli matrices, and $\lambda_{3}$ is the trilinear coupling that mixes the leptoquarks in this case. The Yukawa Lagrangian for $S_3$ reads
%%%%%%%%%%%%%%
\begin{align}
\label{eq:S3-Yuk}
\mathcal{L}^{S_3}_\mathrm{Yuk.}&=y_{3L}^{ij}\, \overline{Q^C_i} i\tau_2 (\tau^a \, S_3^{a}) L_j +\mathrm{h.c.}\,,
\end{align}
%%%%%%%%%%%%%%
where $y_{3L}$ is the $S_3$ leptoquark Yukawa and the fermion fields are defined as before, in the interaction basis, cf.~discussion below.

\subsubsection{From interaction to mass basis} 
\label{sec:leptonic-mixing}

We define the unitary transformations from the interaction to the mass basis as follows,
%%%%%%%%%%%%%%
\begin{align}
\label{eq:fermion-rotation}
    f_L \to U_{f_L} f_L\,, \qquad\qquad f_R \to U_{f_R} f_R\,,
\end{align}
%%%%%%%%%%%%%%
\noindent where $f\in \lbrace \nu, \ell, d, u\rbrace$ denotes any of the SM fermion fields, i.e., neutrinos ($\nu$), charged leptons ($\ell$), down-quarks ($d$) and up-quarks ($u$).  The left-handed rotations are related through the Cabibbo–Kobayashi-Maskawa (CKM) and Pontecorvo–Maki–Naka\-gawa–Sakata (PMNS) matrices,
%%%%%%%%%%%%%%
\begin{align}
V_{\mathrm{CKM}}&=U_{u_L}^\dagger U_{d_L}\,,\qquad\qquad U_{\mathrm{PMNS}}=U_{\ell_L}^\dagger U_{\nu_L}\,,
\end{align}
%%%%%%%%%%%%%%
for which we use the shorthand notation $V\equiv V_{\mathrm{CKM}}$ and $U\equiv U_{\mathrm{PMNS}}$ in what follows. In our scenario, the $U_{\nu_L}$ matrix is entirely fixed, for a given choice of leptoquark couplings, after diagonalizing the neutrino mass matrix induced at one loop, as will be shown in Sec.~\ref{ssec:nu-masses}. Since the PMNS matrix is known from experiment~\cite{Esteban:2020cvm}, we can then determine $U_{\ell_L}=U_{\nu_L} U^\dagger$ as well, up to unknown Majorana phases. 

After the rotation of the fermions to the mass basis defined in Eq.~\eqref{eq:fermion-rotation}, we can write the leptoquark Yukawa interactions as follows,
%%%%%%%%%%%%%%
\begin{align}
\label{eq:R2t-Yuk-bis}
\mathcal{L}^{\widetilde{R}_2}_\mathrm{Yuk.} &= -\big{(}y_{2L} U_{\ell_L}\big{)}_{ij} \,\bar{d}_{Ri} \ell_{Lj} \widetilde{R}_2^{(2/3)}+\big{(}y_{2L} U_{\nu_L}\big{)}_{ij} \,\bar{d}_{Ri} \nu_{Lj} \widetilde{R}_2^{(-1/3)} + \mathrm{h.c.}\,, \\[0.5em]
\label{eq:S1-Yuk-bis}
\mathcal{L}^{{S}_1}_\mathrm{Yuk.} &= -\big{(}y_{1L} U_{\nu_L}\big{)}_{ij} \,\bar{d}_{Li}^C \nu_{Lj} {S}_1^{(1/3)}+\big{(}V^\ast y_{1L} U_{\ell_L} \big{)}_{ij} \,\bar{u}_{Li}^C \ell_{Lj} {S}_1^{(1/3)} + y_{1R}^{ij}\, \bar{u}_{Ri}^C S_1^{(1/3)} \ell_R^j + \mathrm{h.c.} \,,  \\[0.5em]
\mathcal{L}^{{S}_3}_\mathrm{Yuk.} &= -\big{(}y_{3L} U_{\nu_L}\big{)}_{ij} \,\bar{d}_{Li}^C \nu_{Lj} {S}_3^{(1/3)} -\sqrt{2}\big{(}y_{3L} U_{\ell_L}\big{)}_{ij} \,\bar{d}_{Li}^C \ell_{Lj} {S}_3^{(4/3)}  \nonumber\\[0.3em]
\label{eq:S3-Yuk-bis}
&+ \sqrt{2}\big{(}V^\ast y_{3L} U_{\nu_L}\big{)}_{ij} \, \bar{u}_{Li}^C  \nu_{Lj}S_3^{(-2/3)} - \big{(}V^\ast y_{3L} U_{\ell_L}\big{)}_{ij} \, \bar{u}_{Li}^C  \ell_{Lj}S_3^{(1/3)}+  \mathrm{h.c.} \,, 
\end{align}
%%%%%%%%%%%%%%
 where we work in the basis where down-quark Yukawas are diagonal (i.e., $U_{d_L}=\mathbb{1}$), so that the CKM matrix will only appear in the couplings to left-handed up-quarks (i.e., $U_{u_L}=V^\dagger$). Furthermore, we assume that right-handed fields are aligned in the mass basis (i.e.,~$U_{d_R}=U_{u_R}=U_{\ell_R}=\mathbb{1}$), as they do not play an important role in our study. For convenience, we define the primed Yukawa couplings,
%%%%%%%%%%%%%
\begin{align}
\label{eq:primed-yuk-S1}
y_{1L}^\prime \equiv y_{1L} U_{\ell_L}\,,\qquad\quad  y_{2L}^\prime \equiv y_{2L} U_{\ell_L}\,,\qquad\quad y_{3L}^\prime \equiv y_{3L} U_{\ell_L}\,.
\end{align}
%%%%%%%%%%%%%

\noindent which are the leptoquark couplings directly entering physical observables in our setup. 

\subsection{Radiative Neutrino Masses} 
\label{ssec:nu-masses}

\begin{figure}[!t]
\begin{center}
\includegraphics[width=0.88\textwidth]{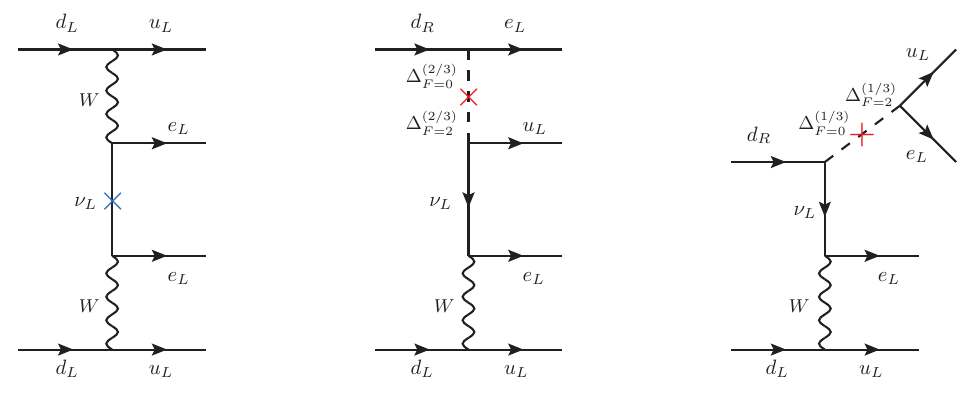} 
\end{center}
\caption{\sl\small Leading contributions to $0\nu\beta\beta$ from the standard $d=5$ operators (left panel), which is induced at one-loop (cf.~Fig.~\ref{fig:diag-loop}), and from the $d=6$ low-energy effective operators (center and right panels).
\label{fig:0nubb}}
\end{figure}

After electroweak symmetry breaking, the leptoquark states with electric charge $Q=-1/3$ will mix with each other, inducing nonzero neutrino Majorana masses via the one-loop diagrams depicted in Fig.~\ref{fig:diag-loop}~\cite{AristizabalSierra:2007nf}. We denote the mass eigenstates by ${S}_\pm^{(-1/3)}$, which are related to the fields in the interaction basis by
%%%%%%%%%%%%%%%%
\begin{equation}
   \begin{pmatrix}
 S_{1(3)}^{(-1/3)\,\ast} \\ \widetilde{R}_2^{(-1/3) } 
\end{pmatrix}
=
\begin{pmatrix}
 \cos\theta & \sin\theta  \\[0.4em] 
-\sin\theta & \cos\theta  
\end{pmatrix}
 \begin{pmatrix}
 {S_+} \\[0.4em] {S_-} 
\end{pmatrix}\,,
\end{equation}
%%%%%%%%%%%%%%%%
\noindent where $v =(\sqrt{2}G_F)^{-1/2}\simeq 246$ GeV is the SM Higgs vacuum expectation value (vev), $G_F$ is the Fermi constant, and the mixing angle $\theta$ satisfies 
%%%%%%%%%%%%%%
\begin{align}
\label{eq:mixing}
\tan 2\theta =  \dfrac{\sqrt{2} \,\lambda_{1(3)} \, v}{m_{S_{1(3)}}^2 - m_{\widetilde{R}_2}^2}\,.
\end{align}
%%%%%%%%%%%%%%
The masses of the physical eigenstates are given by
%%%%%%%%%%%%%%
\begin{align}
    m_{S_\pm}^2 = \frac{1}{2}\left(m_{\widetilde{R}_2}^2 + m_{S_{1(3)}}^2 \pm (m_{S_{1(3)}}^2 - m_{\widetilde{R}_2}^2) \sec(2\theta) \right)\,.
\end{align}
%%%%%%%%%%%%%%
Note, in particular, that the maximal mixing ($\theta=\pi/4$) is obtained for $m_{S_{1(3)}}=m_{\tilde{R}_2}$.~\footnote{The leptoquark masses are constrained by the searches of pair-produced leptoquarks at the LHC. The current limits imply that their masses must lie above $\approx 1.5~\mathrm{TeV}$, with the exact value depending on the Yukawa couplings, cf.~e.g.~Ref.~\cite{Angelescu:2021lln}.} The mixing of the two leptoquarks induces contributions to the Majorana neutrino mass matrix ($m_\nu$),
%%%%%%%%%%%%%%
\begin{align}
\label{eq:nu-mass}
\mathcal{L}_{\nu}\supset -\dfrac{1}{2} (m_\nu)_{ij}\, \overline{\nu_{L\,i}^C} \nu_{L\,j}\,,
\end{align}
%%%%%%%%%%%%%%
\noindent which is given by
%%%%%%%%%%%%%%
\begin{align}
\label{eq:mnu}
m_\nu =  - \dfrac{3 }{16 \pi^ 2} \dfrac{v}{\sqrt{2}}  \sin 2\theta \log \dfrac{m_{S_+}}{m_{S_-}}  \Big{(} y_{2L}^T\cdot \hat{y}_d \cdot y_{L} +y_{L}^T\cdot \hat{y}_d \cdot y_{2L}\Big{)}   \,,
\end{align}
%%%%%%%%%%%%%%
where $\hat{y}_d = \mathrm{diag}(y_d,y_s,y_b)$ denotes the SM down-quark Yukawa matrix, and $y_L$ must be  replaced by $y_{1L}$ ($y_{3L}$), respectively, for the scenario where $\widetilde{R}_2$ mixes with $S_1$ ($S_3$).~\footnote{Notice that the summation over quark flavors in Eq.~\eqref{eq:mnu} is accounted for by the matrix multiplication.} In the decoupling limit with degenerate masses $M$, i.e.,
$m_{\widetilde{R}_2} \sim m_{S_{1(3)}} \sim M \gg v $
, we can show that $m_\nu$ reduces to
%%%%%%%%%%%%%%
\begin{align}
\label{eq:mnus}
m_\nu \simeq  - \dfrac{3 \lambda}{32 \pi^2} \dfrac{v^2}{M^2} \,\Big{(} y_{2L}^T\cdot \hat{y}_d \cdot y_{L} +y_{L}^T\cdot \hat{y}_d \cdot y_{2L}\Big{)}\,,
\end{align}
%%%%%%%%%%%%%%
 where $\lambda$ denotes $\lambda_1$ ($\lambda_3$) for the scenario with the singlet (triplet) leptoquark. The unitary matrices $U_{\nu_L}$ can be determined by diagonalizing the neutrino mass matrix or a fixed choice of $y_L$ and $y_{2L}$. Besides generating neutrino masses, leptoquarks should also contribute to other LNV observables such as $0\nu\beta\beta$ decays. These contributions can be described via $SU(3)_c\times SU(2)_L\times U(1)_Y$ invariant operators with odd-dimensions~\cite{Kobach:2016ami}. They start at $d=7$ and can be relevant since leptoquarks directly contribute at tree level to $0\nu\beta\beta$, with effects that can be chirality enhanced, as illustrated in Fig.~\ref{fig:0nubb}.

A relevant question at this point is the maximal leptoquark mass that can successfully explain the observed neutrino masses. To roughly estimate this value, we can assume that neutrino masses are mainly generated by couplings to a single quark-flavor $q$ (e.g.,~the $b$-quark), and replace $\lambda/M = \mathcal{O}(1)$ in Eq.~\eqref{eq:mnu},~\footnote{This combination of parameters can be probed by the oblique corrections, as will be shown in Sec.~\ref{sec:pheno}.} allowing us to write
%%%%%%%%%%%%%%
\begin{align}
\label{eq:nda}
m_\nu \simeq 0.1~\mathrm{eV} \,\bigg{[}\dfrac{|\lambda|}{M}\bigg{]}\bigg{[}\dfrac{m_q}{m_b}\bigg{]}\bigg{[}\dfrac{M}{10^8~\mathrm{TeV}}\bigg{]}\bigg{[}\dfrac{|y_{2L} y_L|/M^2}{(10^8~\mathrm{TeV})^{-2}}\bigg{]}\,,
\end{align}
%%%%%%%%%%%%%%
 
\noindent where $m_q \in \lbrace m_d,m_s,m_b \rbrace$ denotes the dominant quark-mass that is generating neutrino masses in Fig.~\ref{fig:diag-loop}, and we omit the flavor indices of the leptoquark Yukawas. We find that leptoquark masses as large as $(10^8 - 10^{10})$ TeV can generate viable neutrino masses for perturbative leptoquark Yukawas. 

From Eq.~\eqref{eq:nda}, we can also determine the dependence of neutrino masses on $M$ in the decoupling limit (i.e.,~for $M\gg v$), as well as its interplay with low-energy meson decays probing the leptoquark Yukawas. The leading leptoquark contributions to the latter arise through $d=6$ operators, with effective coefficient proportional to $|y_{\mathrm{LQ}}|^2/M^2$, where $y_{\mathrm{LQ}}$ denotes a generic leptoquark coupling. Therefore, for fixed neutrino masses, we find that large values of $|y_{2L}|^2/M^2$ and $|y_{L}|^2/M^2$ are only allowed for small leptoquark masses $M$, since $|\lambda|/M$ is bounded by~electroweak observables (cf.~Sec.~\ref{sec:pheno}). In other words, this leptoquark mechanism of neutrino mass can only be tested in low-energy flavor observables for sufficiently small values of $M$ (cf.~Sec.~\ref{sec:pheno}). 
This conclusion also holds for $0\nu\beta\beta$ decays, since the leptoquark contributions $\propto | \lambda\, y_{2L}\, y_L|/M^4$, as it will be quantified in Sec.~\ref{sec:0nubb}. Finally, note that for large $M$ values, it is impossible to distinguish with current experiments these leptoquark scenarios from other models for neutrino masses such as the type-I seesaw~\cite{Minkowski:1977sc}.

%%%%%%%%%%%%%%
%%%%%%%%%%%%%%

%%%%%%%%%%%%%%%%%%%%%%%%%%%%%%%%%%%%%%%%
%%%%%%%%%%%%%%%%%%%%%%%%%%%%%%%%%%%%%%%%
\section{EFT Approach for $0\nu\beta\beta$}\label{sec:0nubb}
%%%%%%%%%%%%%%%%%%%%%%%%%%%%%%%%%%%%%%%%
%%%%%%%%%%%%%%%%%%%%%%%%%%%%%%%%%%%%%%%%

New physics contributions to $0\nu\beta\beta$ can be parameterized in general by means of a low-energy Effective Field Theory (EFT)~\cite{Cirigliano:2018yza}. The leading effects are typically the ones from the $d=5$ Weinberg operator that induces nonzero neutrino masses (Eq.~\eqref{eq:nu-mass}). However, in concrete models, there may be additional contributions that can overcome  the ones induced by neutrino masses, despite the naive predictions from the EFT power counting. In the following, we first describe $0\nu\beta\beta$ decays in terms of a low-energy EFT with $d\leq 6$ operators, which is then matched, at the electroweak scale, to the EFT invariant under the full SM gauge symmetry group, $SU(3)_c\times SU(2)_L\times U(1)_Y$, namely the SMEFT~\cite{Grzadkowski:2010es}.

\subsection{Low-energy EFT} In terms of an effective Lagrangian invariant under $SU(3)_c\times U(1)_\mathrm{em}$, the leading LNV operators beyond neutrino masses appear at $d=6$,
%%%%%%%%%%%%%
\begin{align}
\label{eq:left}
\mathcal{L}^{(6)}_\mathrm{eff.} = \sqrt{2}\, G_F
 \sum_a C_a^{(6)}\, O_a^{(6)} + \mathrm{h.c.}\,,
\end{align}
%%%%%%%%%%%%%

\noindent where $C_a$ stand for the effective coefficients related to the effective operators $O_a^{(6)}$,
%%%%%%%%%%%%%
\begin{align}
%%%%%%%%%%%%%
\big{[}O_{V_L}^{(6)}\big{]}_{ij} &= \big{(}\bar{u}_L \gamma^\mu d_L\big{)}\big{(} \bar{\ell}_{Ri} \gamma_\mu \nu_{Lj}^C\big{)}\,, 
%%%%%%%%%%%%%%
&\big{[}O_{V_R}^{(6)}\big{]}_{ij} &= \big{(}\bar{u}_R \gamma^\mu d_R\big{)}\big{(} \bar{\ell}_{Ri} \gamma_\mu \nu_{Lj}^C\big{)}\,,\nonumber\\*[0.3em]
%%%%%%%%%%%%%%
%%%%%%%%%%%%%%
\big{[}O_{S_L}^{(6)}\big{]}_{ij} &= \big{(}\bar{u}_R  d_L\big{)}\big{(} \bar{\ell}_{Li}  \nu_{Lj}^C\big{)}\,, 
%%%%%%%%%%%%%
&\big{[}O_{S_R}^{(6)}\big{]}_{ij} &= \big{(}\bar{u}_L d_R\big{)}\big{(} \bar{\ell}_{Li}  \nu_{Lj}^C\big{)}\,,\\*[0.3em]
%%%%%%%%%%%%%
\big{[}O_{T}^{(6)}\big{]}_{ij} &= \big{(}\bar{u}_L \sigma^{\mu\nu} d_R\big{)}\big{(} \bar{\ell}_{Ri} \sigma^{\mu\nu}  \nu_{Lj}^C\big{)}\,,\nonumber
\end{align}
%%%%%%%%%%%%%

\noindent where $i,j$ denote lepton flavor indices, and $u(d)$ are the up (down) quark. The above Lagrangian is defined in the physical (mass) basis for charged leptons, whereas the choice of neutrino basis is such that the charged-current interaction is diagonal. The leading contributions to these coefficients appear then at $d=7$ in the SMEFT, via the tree-level exchange of leptoquarks depicted in Fig.~\ref{fig:0nubb}, which gives $C_a^{(6)}=\mathcal{O}(v^3/\Lambda^3)$~\cite{Cirigliano:2017djv}, cf. Sec.~\ref{sec:SMEFT}. The scalar and tensor operators evolve under the QCD Renormalization Group Evolution (RGE), which can be expressed at one-loop as~\cite{Cirigliano:2018yza},
%%%%%%%%%%%%%
\begin{align}
\begin{split}
    \dfrac{\mathrm{d} C_{S_{L(R)}}^{(6)}}{\mathrm{d} \log \mu} &= -6 \, C_F \dfrac{\alpha_s(\mu)}{4\pi}C_{S_{L(R)}}^{(6)}(\mu)\,, \qquad\quad
\dfrac{\mathrm{d} C_{T}^{(6)}}{\mathrm{d} \log \mu} = 2 \, C_F \dfrac{\alpha_s(\mu)}{4\pi}C_{T}^{(6)}(\mu)\,,
\end{split}
\end{align}
%%%%%%%%%%%%%

\noindent where $C_F=(N_c^2-1)/(2 N_c)$, $N_c=3$ is the number of colors and $\alpha_s(\mu)$ is the strong structure constant at the scale $\mu$. In our analysis, we consider the three-loop running effects from Ref.~\cite{Gracey:2000am} that are summarized in Ref.~\cite{Becirevic:2024pni} (see also Ref.~\cite{Gonzalez-Alonso:2017iyc}). Notice that the vector coefficients are not renormalized by QCD.

In addition to the low-energy $d=6$ operators that are listed above, leptoquarks can also induce $d=7$ operators via the insertion of covariant derivatives~\cite{Cirigliano:2018yza}, or $d=9$ contact operators made of the product of three fermion-bilinears~\cite{Graesser:2016bpz}.~\footnote{Note, in particular, that $d=9$ operators arise in the scenario with the weak-triplet leptoquark $S_3$ via the triple vertex $\smash{W^+S_3^{(-2/3)}S_3^{(-1/3)}}$.} These operators appear at higher orders in the SMEFT, thus being irrelevant to our phenomenological analysis due to the additional $v/\Lambda$ suppression factors.

\subsection{SMEFT description} 
\label{sec:SMEFT}
The complete matching between the low-energy EFT defined in Eq.~(\ref{eq:left}) with its $SU(3)_c\times SU(2)_L \times U(1)_Y$ invariant version is given~in~Ref.~\cite{Cirigliano:2017djv}. There are only a few of these coefficients that arise from the leptoquarks models described in Sec.~\ref{sec:lq-model}. We define the SMEFT Lagrangian as follows,
%%%%%%%%%%%%%
\begin{align}
\mathcal{L}_\mathrm{SMEFT} = \sum_{d\geq 5} \sum_a \dfrac{\mathcal{C}_a^{(d)}}{\Lambda^{d-4}}\, \mathcal{O}_a^{(d)} \,,
\end{align}
%%%%%%%%%%%%%
where $\mathcal{C}_a^{(d)}$ are effective coefficients of $d$-dimensional operators, and $\Lambda$ denotes the EFT cutoff. The lowest-order operators that are relevant to us appear at $d=7$,
%%%%%%%%%%%%%
\begin{align}
 \big{[} \mathcal{O}_{LLQ\overline{d}H}^{(1)}\big{]}_{ijkl} &= \epsilon_{ab} \epsilon_{mn}\left(\overline{d_{Rk}} L_i^a\right) \left(\overline{L_j^{m\,C}} Q_l^{b}\right) H^n\, ,\\[0.35em]
 \big{[}\mathcal{O}_{LLQ\overline{d}H}^{(2)}\big{]}_{ijkl} &= \epsilon_{am} \epsilon_{bn}\left(\overline{d_{Rk}} L_i^a\right) \left(\overline{L_j^{m\,C}} Q_l^{b} \right) H^n \, ,\\[0.35em]
 \big{[}\mathcal{O}_{Leu\overline{d}H}\big{]}_{ijkl} &= \epsilon_{bn} \left(\overline{d_{Rk}} \gamma_\mu u_{Rl}\right) \left(\overline{L_i^{b\,C}}\, \gamma^\mu e_{Rj}\right) H^n \, ,
\end{align}
%%%%%%%%%%%%%

\noindent where $i,j,k,l$ are flavor indices, and we explicitly write the $SU(2)_L$ indices, with $\epsilon_{12}=-\epsilon_{21}=+1$, adopting the same conventions of Ref.~\cite{Lehman:2014jma}. In the scenario with $\widetilde{R}_2$ and $S_1$ defined in Eq.~\eqref{eq:lag-R2t-S1}, the effective coefficients at the matching scale $\mu=\Lambda$ read
%%%%%%%%%%%%%
\begin{align}
\label{eq:matching-R2-S1}
\begin{split}
\dfrac{1}{\Lambda^3}\big{[} \mathcal{C}_{LLQ\overline{d}H}^{(1)}\big{]}_{ijkl} &=-\dfrac{1}{\Lambda^3}
\big{[} \mathcal{C}_{LLQ\overline{d}H}^{(2)}\big{]}_{ijkl} =\dfrac{\lambda_1}{m_{S_1}^2 m_{\widetilde R_2}^2} { y_{1L}^{lj} \,y_{2L}^{ki}} \,,\\[0.35em]
\dfrac{1}{\Lambda^3}\big{[} \mathcal{C}_{Leu\overline{d}H}\big{]}_{ijkl} &= \dfrac{1}{2}\dfrac{\lambda_1}{m_{S_1}^2 m_{\widetilde R_2}^2} { y_{1R}^{lj} \,y_{2L}^{ki}} \,.
\end{split}
\end{align}
%%%%%%%%%%%%%

\noindent Similarly, for the scenario with $\widetilde{R}_2$ and $S_3$ defined in Eq.~\eqref{eq:lag-R2t-S3}, the effective coefficients are given by
%%%%%%%%%%%%%
\begin{align}
\label{eq:matching-R2-S3}
\dfrac{1}{\Lambda^3}\big{[} \mathcal{C}_{LLQ\overline{d}H}^{(1)}\big{]}_{ijkl}= \dfrac{1}{\Lambda^3}\big{[} \mathcal{C}_{LLQ\overline{d}H}^{(2)}\big{]}_{ijkl} &= - \dfrac{\lambda_3}{m_{S_3}^2 m_{\widetilde R_2}^2} { y_{3L}^{lj} \,y_{2L}^{ki}}\,.
\end{align}
%%%%%%%%%%%%%

\noindent The operators listed above can be matched at tree level to the low-energy EFT at the $\mu=\mu_\mathrm{ew}$ scale,
%%%%%%%%%%%%%
\begin{small}
\begin{align}
\label{eq:smeft-matching-1}
\big{[}C_{V_R}^{(6)}\big{]}_{ij} & = \dfrac{1}{\sqrt{2}} \dfrac{v^3}{\Lambda^3} \sum_{k} (U_{\ell_L}^\ast)_{k j} \, \big{[} \mathcal{C}_{Leu\overline{d}H}\big{]}_{k i 11}^\ast \,,\\[0.3em]
\label{eq:smeft-matching-2}
\big{[}C_{S_R}^{(6)}\big{]}_{i j} & =\dfrac{1}{2\sqrt{2}} \dfrac{v^3}{\Lambda^3}
\sum_{k l m} (U_{\ell_L}^\ast)_{k i} (U_{\ell_L}^\ast)_{m j}(V)_{1 l} \Big{(}\big{[}\mathcal{C}_{LLQ\bar{d}H}^{(2)}\big{]}_{k m 1l}^\ast-\big{[}\mathcal{C}_{LLQ\bar{d}H}^{(2)}\big{]}_{m k 1l}^\ast+\big{[}\mathcal{C}_{LLQ\bar{d}H}^{(1)}\big{]}_{k m 1l}^\ast\Big{)}\,,\\[0.3em]
\label{eq:smeft-matching-3}
\big{[}C_{T}^{(6)}\big{]}_{i j} & =\dfrac{1}{8\sqrt{2}}\dfrac{v^3}{\Lambda^3}\sum_{k l m} (U_{\ell_L}^\ast)_{k i} (U_{\ell_L}^\ast)_{m j}(V)_{1 l} \Big{(}\big{[}\mathcal{C}_{LLQ\bar{d}H}^{(2)}\big{]}_{k m 1l}^\ast+\big{[}\mathcal{C}_{LLQ\bar{d}H}^{(2)}\big{]}_{m k 1l}^\ast+\big{[}\mathcal{C}_{LLQ\bar{d}H}^{(1)}\big{]}_{k m 1l}^\ast\Big{)}\,,
\end{align}
\end{small}
%%%%%%%%%%%%%

\noindent which is expressed in the basis introduced in Eq.~\eqref{eq:left}, by means of the fermion rotations defined in Eq.~\eqref{eq:fermion-rotation}. From the matching to the SMEFT of the two leptoquark models introduced in Eqs.~\eqref{eq:matching-R2-S1} and \eqref{eq:matching-R2-S3}, we see that the contributions to the $0\nu\beta\beta$ will be different depending if $\widetilde{R}_2$ mixes with a singlet $(S_1)$ or with a triplet $(S_3)$ leptoquark to generate neutrino masses, as it could be anticipated from the chiralities of the fermions in Fig.~\ref{fig:0nubb}. 

\subsection{$0\nu\beta\beta$ decays}

In this paper, we consider the EFT expressions for the $0\nu\beta\beta$ half-life derived in full generality in Ref.~\cite{Cirigliano:2018yza}, which have been recently automatized in Ref.~\cite{Scholer:2023bnn}. These expressions are obtained by using the Chiral Lagrangian~\cite{Cirigliano:2017djv,Cirigliano:2017tvr,Cirigliano:2018hja}, which is matched onto the $SU(3)_c\times U(1)_\mathrm{em}$ EFT describing $0\nu\beta\beta$ decays, and combined with the nuclear matrix elements that have been estimated in Ref.~\cite{Menendez:2017fdf} (see also Ref.~\cite{Hyvarinen:2015bda,Barea:2015kwa}). The master formula for the inverse half-life reads~\cite{Cirigliano:2018yza} (see also Ref.~\cite{Doi:1985dx}),
%%%%%%%%%%%%%
\begin{align}
\begin{split}
\left(T_{1/2}^{0\nu}\right)^{-1}   = g_A^4 \,&\Big{\lbrace} G_{01}\,|\mathcal{A}_\nu|^2 +4 \, G_{02}\,|\mathcal{A}_E|^2 + 2\,  G_{04}\left[|\mathcal{A}_{m_e}|^2+\mathrm{Re}(\mathcal{A}_{m_e}^\ast {\cal A}_\nu)\right]+ G_{09}\,|\mathcal{A}_M|^2\\
&-2 \, G_{03}\,\mathrm{Re}(\mathcal{A}_\nu \mathcal{A}_E^\ast + 2 \mathcal{A}_{m_e} {\cal {A}}_E^\ast)+G_{06}\,\mathrm{Re}(\mathcal{A}_\nu {\cal {A}}_M^\ast)\Big{\rbrace}\,,
\end{split}
\end{align}
%%%%%%%%%%%%%
where $G_{0k}$ are phase-space factors reported in Ref.~\cite{Horoi:2017gmj}, $g_A$ is the nucleon axial charge,
and $\mathcal{A}_a$ are sub-amplitudes which can be decomposed as~\cite{Cirigliano:2018yza}
%%%%%%%%%%%%%
\begin{align}
\cal{A}_\nu &= \dfrac{m_{\beta\beta}}{m_e} \mathcal{M}_{\nu}^{(3)} + \dfrac{m_N}{m_e} \mathcal{M}_{\nu}^{(6)} \,, \\[0.35em]
\mathcal{A}_{M} &= \dfrac{m_N}{m_e} \mathcal{M}_M^{(6)}\,, \\[0.35em]
{\cal A}_E &= \mathcal{M}_{E,L}^{(6)} +\mathcal{M}_{E,R}^{(6)} \,, \\[0.35em]
{\cal{A}}_{m_e} &= \mathcal{M}_{m_e,L}^{(6)} +\mathcal{M}_{m_e,R}^{(6)}\,,  
\end{align}
%%%%%%%%%%%%%
where $m_{\beta\beta} \equiv \sum_i m_{\nu_i} \, U_{ei}^2$ is the electron-neutrino effective Majorana mass, $m_N$ denotes the nucleon mass, $m_e$ is the electron mass and we have kept operators up to $d=6$. The contributions induced by light Majorana neutrinos are encapsulated in~\cite{Cirigliano:2018yza}  
%%%%%%%%%%%%%
\begin{align}
\label{eq:NME3}
\mathcal{M}_{\nu}^{(3)}=-V_{ud}^2\bigg{(}-\dfrac{1}{g_A^2}{M}_F + {M}_{GT} + {M}_T + \dfrac{2 m_\pi^2 g_{\nu}^{NN}}{g_A^2} M_{F,sd}\bigg{)}\,,
\end{align}
%%%%%%%%%%%%%

\noindent where $M_F$, $M_{GT}$ and $M_T$ are the Fermi (F), Gamow-Teller (GT) and Tensor (T) long-range Nuclear Matrix-Elements (NMEs), respectively, which are estimated e.g.~in Ref.~\cite{Menendez:2017fdf}. The short-range NME $M_{F,sd}$ in this expression arises from the hard-neutrino exchange, and it depends on the low-energy constant $g_\nu^{NN} = \mathcal{O}(F_\pi^{-2})$~\cite{Cirigliano:2018hja}. The contributions from the $d=6$ operators appear instead in the sub-amplitudes~\cite{Cirigliano:2018yza},
%%%%%%%%%%%%%
\begin{align}
\mathcal{M}_\nu^{(6)} &= V_{ud}\, \bigg{[} \dfrac{B}{m_N}(C_{S_L}^{(6)}-C_{S_R}^{(6)}) M_{PS} +C_{T}^{(6)} M_{T6} \bigg{]} \,,\\[0.35em]
\mathcal{M}_M^{(6)} &= V_{ud}\, C_{V_L}^{(6)}M_M\,,\\[0.35em]
\mathcal{M}_{m_e,L(R)}^{(6)} &= V_{ud} \,C_{V_{L(R)}}^{(6)} M_{m_e,L(R)}\,,\\[0.35em]
\mathcal{M}_{E,L(R)}^{(6)} &= V_{ud} \, C_{V_{L(R)}}^{(6)} M_{E,L(R)}\,,
\label{eq:NME6}
\end{align}
%%%%%%%%%%%%%

\noindent where $B\equiv - \langle \bar{q}q\rangle/F_\pi^2 \simeq 2.7$~GeV at $\mu=2$~GeV in the $\overline{\mathrm{MS}}$ scheme~\cite{Cirigliano:2018yza}, 
$M_{PS}$, $M_{T6}$, $M_M$, $M_{m_e,L(R)}$ and $M_{E,L(R)}$ are the remaining NME. Notice that we have omitted the flavor indices of the Wilson coefficients, which should refer to $e$ and $\nu_e$. The nuclear matrix elements needed in Eqs.(\ref{eq:NME3})-(\ref{eq:NME6}), as well as the other 
nuclear-physics inputs used in our analysis, are summarized in Appendix~\ref{app:nuclear-inputs}. These input values coincide with the default setting of the package provided in Ref.~\cite{Scholer:2023bnn}.

\paragraph{Numerical significance} By combining the above expressions, we can define the \emph{effective value} of $m_{\beta\beta}^\mathrm{eff}$ that would be extracted from the experimental value of the half-life $T^{0\nu}_{1/2}$ if $d=6$ operators are present in addition to the light Majorana neutrino masses,
%%%%%%%%%%%%%
\begin{align}
\label{eq:mbb_eff}
|m_{\beta\beta}^\mathrm{eff}|^2 \equiv m_{\beta\beta}^2 + \delta m_{\beta\beta}^{2\,(\mathrm{int})}+ \delta m_{\beta\beta}^{2\,(\mathrm{LQ})}\,,
\end{align}
%%%%%%%%%%%%%

\noindent where the linear and quadratic leptoquark contributions are encapsulated in $\smash{\delta m_{\beta\beta}^{2\,(\mathrm{int})}}$ and $\smash{\delta m_{\beta\beta}^{2\,(\mathrm{LQ})}}$, respectively. For the $^{136}\mathrm{Xe}$ isotope, we find that 
%%%%%%%%%%%%%
\begin{align}
\delta m_{\beta\beta}^{2\,(\mathrm{int})} &\simeq - \Big{(}
1.1 \,{\rm meV}\Big{)}^2 \left(\dfrac{10^3\,{\rm TeV}}{\Lambda}\right)^3 \left(\frac{m_{\beta\beta}}{\rm meV}\right) \left( \Bar{C}_{S_R}^{(6)} - 0.7 \,\Bar{C}_{T}^{(6)} - 8 \times 10^{-4}\,\Bar{C}_{V_R}^{(6)}\right)^\ast +{\rm h.c.}\,,\nonumber\\[0.5em]
\delta m_{\beta\beta}^{2\,(\mathrm{LQ})}  &\simeq \Big{(}1.2 \,{\rm meV}\Big{)}^2 \left(\dfrac{10^3\,{\rm TeV}}{\Lambda}\right)^6 \bigg{[}\left(\vert \Bar{C}_{S_R}^{(6)}\vert^2 + 0.6 \,\vert \Bar{C}_{T}^{(6)}\vert^2 + 8 \times 10^{-6} \,\vert \Bar{C}_{V_R}^{(6)}\vert^2\right) \nonumber  \\
    &\quad-2\,\mathrm{Re}\Big{(}\Bar{C}_{S_R}^{(6)}\,\Bar{C}_{T}^{(6)\ast}\Big{)}-1\times 10^{-3}\, \mathrm{Re}\Big{(}\Bar{C}_{V_R}^{(6)}\,\Bar{C}_{T}^{(6)\ast}\Big{)} -2\times 10^{-5}\, \mathrm{Re}\Big{(}\Bar{C}_{V_R}^{(6)}\,\Bar{C}_{S_R}^{(6)\ast} \Big{)}\bigg{]}\,,
\end{align}
%%%%%%%%%%%%%

\noindent where the effective coefficients are evaluated at the scale $\mu=2$~GeV, and we have defined $\Bar{C}_a^{(6)}\equiv C_a^{(6)}\,\Lambda^3/v^3$, where $\Lambda$ denotes the EFT cutoff, since $C_a^{(6)}= \mathcal{O}(v^3/\Lambda^3)$ in the SMEFT. From the above expressions, it is clear that the EFT contributions are not suppressed by the neutrino masses, which are replaced by the nuclear energy scale, $E\approx 100$~MeV, cf.~Fig.~\ref{fig:0nubb}. In particular, the largest EFT contributions are those induced by the scalar and tensor operators~\cite{Cirigliano:2017djv}. By using the current experimental limit $T_{1/2}^{0\nu} > 3.8 \times 10^{26}$ yr (90
\% C.L.) obtained by KamLAND-Zen for $^{136}{\rm Xe}$~\cite{KamLAND-Zen:2024eml}, we find that 
%%%%%%%%%%%%%
\begin{align}
\quad  \vert C_{S_{R,L}}^{(6)} \vert &\lesssim 2.8 \times 10^{-10}\,,   & \vert C_{T}^{(6)} \vert &\lesssim 3.8 \times 10^{-10}\,,\nonumber\\[0.35em]
  \vert C_{V_{L}}^{(6)} \vert &\lesssim 8.1 \times 10^{-10}\,,  & \vert C_{V_{R}}^{(6)} \vert &\lesssim 1.0 \times 10^{-7} \,, \label{eq:limits-EFT-coef}
 % & { \rm all } \; @ & \; \mu = 2\rm \; GeV\,,
\end{align}
%%%%%%%%%%%%%%

\noindent where the renormalization scale is fixed to $\mu=2$~GeV, and the flavor indices are once again omitted. These limits can be improved by a factor 10 by the LEGEND-1000 experiments~\cite{LEGEND:2021bnm}.~\footnote{The upper limits in Eq.~\eqref{eq:limits-EFT-coef}  depend on certain Chiral Lagrangian parameters that have not yet been determined, see Appendix~\ref{app:nuclear-inputs}. For the limits presented in Eq.~\ref{eq:limits-EFT-coef}, we assumed these coefficients to be zero. However, varying them within the range $[-\sqrt{4\pi},\sqrt{4\pi}]$ does not affect the bounds on $\smash{\vert C_{S_{L,R}}^{(6)} \vert}$, while the upper limits on $\smash{\vert C_{T}^{(6)} \vert}$ and $\smash{\vert C_{V_{L,R}}^{(6)} \vert}$ may vary by up to factors of 10 and $10^2$, respectively. Nevertheless, these variations do not impact our numerical results for the leptoquark model, as the most relevant coefficients in this case are $\smash{\vert C_{S_R}^{(6)} \vert}$ and $\smash{\vert C_{T}^{(6)} \vert}$.}

%%%%%%%%%%%%%%
\begin{figure}[!t]
\begin{center}
\includegraphics[width=0.78\textwidth]{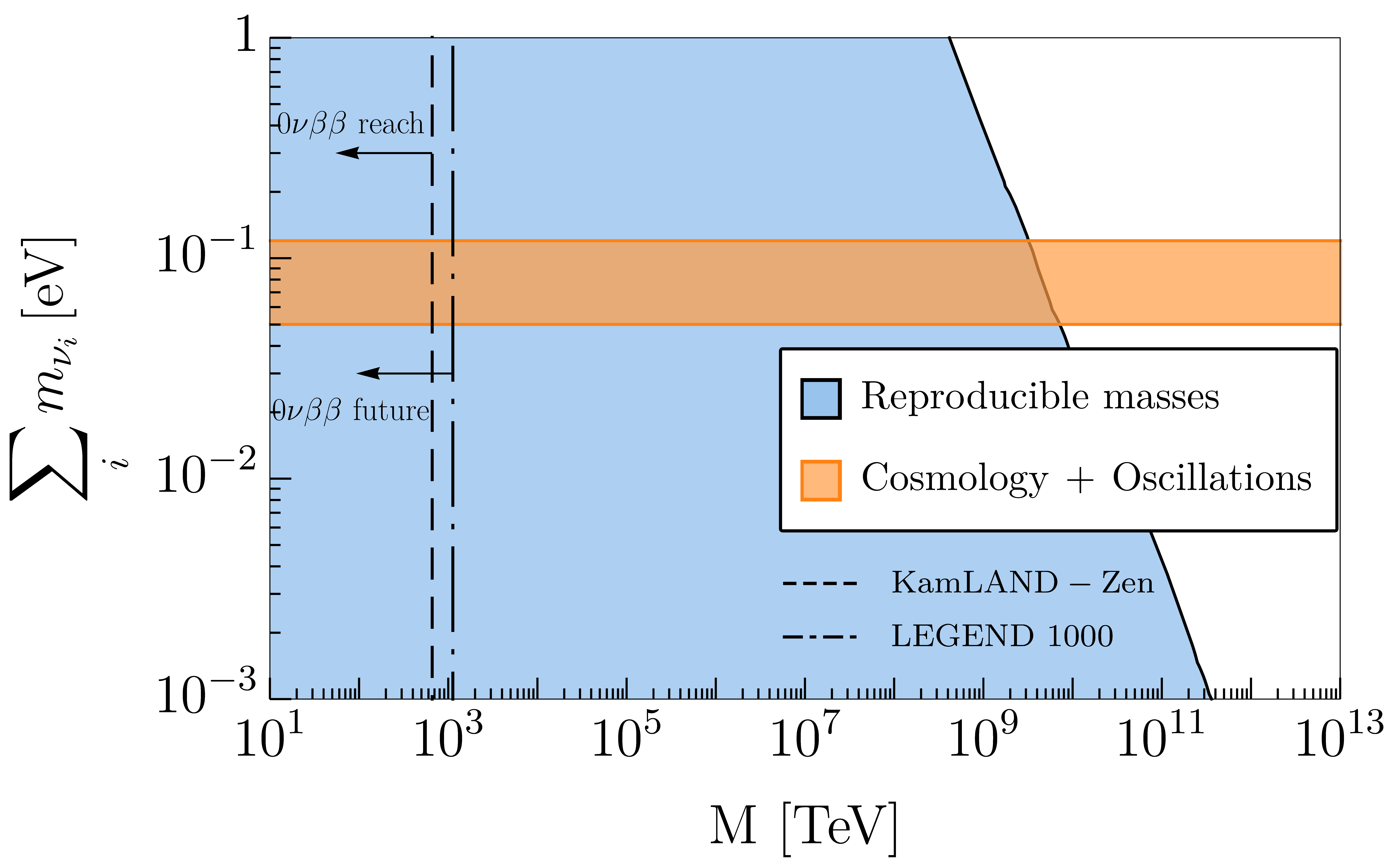} 
\end{center}
\caption{\sl\small Allowed values for the sum of the neutrino masses ($\sum_i {m_{\nu_i}}$) as a function of the leptoquark mass $M$, assuming  the two leptoquarks to be degenerate in mass (blue region). The leptoquark Yukawas are varied within the naive perturbativity premise ($\lesssim \sqrt{4\pi}$) and the trilinear couplings are fixed to $\lambda_{1(3)}=M/2$. The orange region corresponds to neutrino masses compatible with neutrino oscillation data, cf.~Eq.~\eqref{eq:oscillation}, as well as the indirect upper limits from cosmology~\cite{Planck:2018vyg}.}
\label{fig:nu-masses}
\end{figure}
%%%%%%%%%%%%%%

By combining Eqs.~\eqref{eq:matching-R2-S1}--\eqref{eq:matching-R2-S3} with Eqs.~\eqref{eq:smeft-matching-1}--\eqref{eq:smeft-matching-3}, we find that the leptoquark models introduced in Sec.~\ref{sec:lq-model} induce the following combinations of effective coefficients,
%%%%%%%%%%%%%
\begin{align}
\begin{split}
\widetilde{R}_2-S_1: & \qquad C_{S_R} =-4 \, C_T = \dfrac{\lambda_1 v^3 y_{1L}^{\prime 11\,\ast}\,y_{2L}^{\prime 11\,\ast}}{2\sqrt{2}M^4} \,, \qquad C_{V_R} = \dfrac{\lambda_1 v^3\,y_{1R}^{11\,\ast} \,y_{2L}^{\prime 11\,\ast}}{2\sqrt{2}M^4}\,,  \\[0.3em]
\widetilde{R}_2-S_3: & \qquad C_{S_R} =+ \frac{4}{3} \, C_T =  -\dfrac{\lambda_3 v^3 y_{3L}^{\prime 11\,\ast}\, y_{2L}^{\prime 11\,\ast}}{2 \sqrt{2}M^4}\,,
\end{split}
\end{align}
%%%%%%%%%%%%% 
where RGE effects are not yet included.  By using the above expressions, we obtain the following upper limits for the $S_1$-$\widetilde{R}_2$ scenario, 
%%%%%%%%%%%%%
\begin{align}
\label{eq:limits-lq-1}
\dfrac{|\lambda_1 y_{1L}^{\prime 11} y_{2L}^{\prime 11}|}{M^4} &\lesssim 
(351~\mathrm{TeV})^{-3}\,, &\dfrac{|\lambda_1 y_{1R}^{11} y_{2L}^{\prime 11}|}{M^4} &\lesssim (37~\mathrm{TeV})^{-3}\,,
\end{align}
%%%%%%%%%%%%%

\noindent and, similarly, for $S_3$-$\widetilde{R}_2$, 
%%%%%%%%%%%%%
\begin{align}
\label{eq:limits-lq-2}
\dfrac{|\lambda_3 y_{3L}^{\prime 11} y_{2L}^{\prime 11}|}{M^4} &\lesssim (364 ~\mathrm{TeV})^{-3}\,,%& { \rm all } \; @ & \; \mu = 1\rm \; TeV\,,
\end{align}
%%%%%%%%%%%%%

\noindent where we have taken degenerate masses $M \equiv m_{S_{1(3)}}=m_{\widetilde{R}_2}$, and used the primed coefficients defined in Eq.~\eqref{eq:primed-yuk-S1} for the first generation of fermions. The above upper limits include the QCD RGE effects from the $\Lambda\approx 100~\mathrm{TeV}$ down to $\mu=2$~GeV, which amount to a $\mathcal{O}(20\%)$ effect, which is much smaller than potential uncertainties arising, e.g.,~from the estimation of the nuclear matrix elements.~\footnote{We stress once again that varying the unknown low-energy parameters of the Chiral Lagrangian within the $[-\sqrt{4\pi},\sqrt{4\pi}]$ range has a minor impact on these results, which are only changed by a few percent for the combinations involving $|y_{1L}^\prime y_{2L}^\prime|$ and $|y_{3L}^\prime y_{2L}^\prime|$ couplings, and up to $50\%$ for the $|y_{1R} y_{2L}^\prime|$ combination, which is however unrelated to neutrino masses [cf.~Eq.~\eqref{eq:mnus}].}

In Fig.~\ref{fig:nu-masses}, we compare the allowed values for the sum of neutrino masses $\sum_{i} m_{\nu_i}$ as a function of the degenerate leptoquark masses $M$, as derived in Sec.~\ref{sec:lq-model}, with the current and future reach of $0\nu\beta\beta$ decay experiments. Moreover, we superimpose the constraints from oscillation data~\cite{Esteban:2020cvm} and cosmology~\cite{Planck:2018vyg} on the same plot.  The trilinear couplings are fixed to $\lambda_{1(3)} = M/2$ and the Yukawa couplings are varied within the perturbativity range, see the discussion in Sec.~\ref{sec:numerical}. We find that $0\nu\beta\beta$ experiments can in principle probe a substantial part of the parameter space, namely leptoquark masses up to $\mathcal{O}(10^3~\mathrm{TeV})$. Larger masses would remain beyond the reach of these experiments (and other low-energy probes) while still providing viable neutrino masses, for values $M \lesssim 10^9$ TeV.  Note, in particular, that such large masses are only realized in scenarios where the bottom quark is running in the loop depicted in Fig.~\ref{fig:diag-loop}, whereas masses of $\mathcal{O}(10^{6}~\mathrm{TeV})$ are allowed for leptoquarks coupled to first-generation quarks.

%%%%%%%%%%%%%%%%%%%%%%%%%%%%%%%%%%%%%%%%
%%%%%%%%%%%%%%%%%%%%%%%%%%%%%%%%%%%%%%%%
\section{Phenomenology}\label{sec:pheno}
%%%%%%%%%%%%%%%%%%%%%%%%%%%%%%%%%%%%%%%%
%%%%%%%%%%%%%%%%%%%%%%%%%%%%%%%%%%%%%%%%

In this Section, we discuss the most relevant theoretical and phenomenological constraints on the leptoquark interactions that are relevant for $0\nu\beta\beta$ decays:

\paragraph*{Perturbative Unitarity} The trilinear scalar couplings $\lambda_{1(3)}$ can be bounded by perturbative unitarity~\cite{Lee:1977yc}. We consider the $\smash{\widetilde{R}_2^{(1/3)} h \to \widetilde{R}_2^{(1/3)} h}$ and $\smash{\widetilde{R}_2^{(1/3)} S_{1(3)}^{(-1/3)} \to \widetilde{R}_2^{(1/3)} S_{1(3)}^{(-1/3)}}$ scattering, and impose that the partial-wave amplitude $a_0$ with angular momentum $J=0$ satisfies $|\mathrm{Re}(a_0)|<1/2$. By assuming that the leptoquarks have the same mass $M$, we derive the following constraints
%%%%%%%%%%%%%%%%
\begin{equation}
    \label{eq:unit-cons}
    \dfrac{|\lambda_1|}{M} \lesssim  10\,, \qquad\qquad     \dfrac{|\lambda_3|}{M} \lesssim  9.6\,, 
\end{equation}
%%%%%%%%%%%%%%%%
which are dominated by the processes $\widetilde{R}_2^{(1/3)} h \to \widetilde{R}_2^{(1/3)} h$ and $\widetilde{S}_2^{(-2/3)} h \to \widetilde{S}_2^{(-2/3)} h$, respectively.

\paragraph*{Electroweak Precision Data} The mixing of the two leptoquark states via the couplings $\lambda_{1(3)}$ induce custodial-breaking effects at one loop that contribute to the $T$-parameter~\cite{Peskin:1991sw}, see e.g.~Ref.~\cite{Dorsner:2019itg}. With the assumption of degenerate leptoquark states with mass $M \gg v$, we find that 
%%%%%%%%%%%%%%%%
\begin{equation}
    \label{eq:DeltaT}
    \Delta T_{\widetilde{R}_2 \& S_1} = {\dfrac{N_c \,v^2}{192 \pi c_w^2 s_w^2 m_Z^2}\dfrac{\lambda_1^2}{M^2}}+\dots\,,
\end{equation}
%%%%%%%%%%%%%%%%
and 
%%%%%%%%%%%%%%%%
\begin{equation}
    \label{eq:DeltaT}
    \Delta T_{\widetilde{R}_2 \& S_3} = \dfrac{N_c \,v^2 }{64 \pi  c_w^2 s_w^2 m_Z^2}\dfrac{\lambda_3^2}{M^2} +\dots \,.
\end{equation}
%%%%%%%%%%%%%%%%
where the ellipses represent high-order terms in $1/M$, $m_Z$ is the $Z$-boson mass, and we define $c_w \equiv \cos \theta_w$ and $s_w \equiv \sin \theta_w$, where $\theta_w$ is the Weinberg angle. We consider the results of the latest electroweak fit, which gives $\Delta T_\mathrm{exp}=0.05(11)$~\cite{Baak:2012kk}. By imposing this constraint to $2\sigma$ accuracy, we find that 
%%%%%%%%%%%%%%%%
\begin{equation}
    \label{eq:DeltaT-cons}
    \dfrac{|\lambda_1|}{M} \lesssim { 1.1}\,, \qquad\qquad     \dfrac{|\lambda_3|}{M} \lesssim  0.7\,.
\end{equation}
%%%%%%%%%%%%%%%%
Note, in particular, that these constraints are much more stringent than the ones derived from perturbative unitarity in Eq.~\eqref{eq:unit-cons}.

 \paragraph*{Flavor observables} There are only a few flavor observables that are sensitive to leptoquark masses $M$ above $100~\mathrm{TeV}$, as we consider in this study.~\footnote{We consider large leptoquark masses to restrict our phenomenological analysis of flavor processes, as few processes are currently sensitive to such large energy scales. However, notice that lower $M$ values can also produce viable neutrino masses for smaller leptoquark couplings.} In the following, we will focus on the left-handed leptoquark couplings, which are the relevant ones for neutrino masses, and we will explicitly show that selected flavor observables, in the kaon and lepton sectors, can compete with the constraints from $0\nu\beta\beta$ decays:

%%%%%%%%%%%%%%%
\begin{itemize}
     \item[i)] $\mu N\to e N$ and $\mu \to e \gamma$: Leptoquarks can contribute to Lepton Flavor Violating (LFV) observable such as $\mu\to e$ conversion in nuclei ($\mu N\to e N$), as well as $\mu\to e \gamma$ and $\mu\to eee$ decays. In particular, $\mu N\to e N$ receives tree-level contributions from the product of the leptoquark couplings to muons and electrons. Currently, the most stringent current limit was obtained with gold atoms, $\mathcal{B}_{\mu e}^{(\mathrm{Au})}<7\times 10^{-13}$ (90~$\%$ CL)~\cite{ParticleDataGroup:2022pth}, where $\mathcal{B}_{\mu e}^{(N)} \equiv \mathcal{B}(\mu N\to e N)$ denotes the $\mu\to e$ conversion rate normalized by the muon capture rate. By using e.g.~the EFT analysis of spin-independent $\mu N\to e N$ rates from Ref.~\cite{Kitano:2002mt} (see also Ref.~\cite{Kuno:1996kv}), one can derive the following bounds on the couplings to first-generation quarks
%%%%%%%%%%%%%%%
\begin{align}
\label{eq:limits-muNeN}
\begin{split}
   \dfrac{|y^{\prime\,12}_{1L} \, y^{\prime\,11\ast}_{1L}|}{M^2} &< (470~\mathrm{TeV})^{-2}\,,\qquad\qquad
   \dfrac{|y^{\prime\,12}_{3L} \, y^{\prime\,11\ast}_{3L}|}{M^2} < (830~\mathrm{TeV})^{-2}\,,\\[0.45em]
   \dfrac{|y^{\prime\,12}_{2L} \, y^{\prime\,11\ast}_{2L}|}{M^2} &< (500~\mathrm{TeV})^{-2}\,,
\end{split}
\end{align}
%%%%%%%%%%%%%%%    
where the contributions from the $d=6$ operators induced by each leptoquark state were separately considered, thus neglecting the sub-leading effects from the mixing. These processes will be discussed in detail in Sec.~\ref{ssec:flavor-probes}. Another potentially important constraint is $\mathcal{B}(\mu\to e\gamma)^\mathrm{exp}<4.2\times 10^{-13}$ (90~$\%$ CL.)~\cite{ParticleDataGroup:2022pth}. However, this process is only induced at loop-level in these scenarios, thus not providing meaningful constraints for the leptoquark mass ranges that we consider in this study~\cite{Lavoura:2003xp}.~\footnote{The product of left- and right-handed couplings to the top quark can be severely constrained by $\mu\to e \gamma$ due to the chirality enhancement of the dipoles ($\smash{\propto m_t/m_\mu}$), see e.g.~Ref.~\cite{Feruglio:2018fxo}. However, the right-handed couplings are neglected in our analysis since they do not contribute at leading order to $m_\nu$ and $0\nu\beta\beta$ decays.} 

     \item[ii)] $\Delta F=1$: Another stringent bound on the leptoquark couplings arises from $\Delta F=1$ processes such as $K^+\to \pi^+ \nu\bar{\nu}$, which receive tree-level contributions from leptoquarks coupled to neutrinos with all flavors~\cite{Bobeth:2017ecx,Fajfer:2018bfj}. The experimental determination $\mathcal{B}(K^+\to\pi^+\nu\bar{\nu})^\mathrm{exp}=\big{(}1.14_{-0.33}^{+0.40}\big{)}\times 10^{-10}$ by NA62 is compatible with $\mathcal{B}(K^+\to\pi^+\nu\bar{\nu})^\mathrm{SM}=(7.7\pm 0.6)\times 10^{-11}$~\cite{Brod:2021hsj},~\footnote{See Ref.~\cite{Buras:2021nns} for a detailed discussion on the impact of a different choice of CKM inputs in these SM predictions, in particular, given the long-standing discrepancy in the determination of $|V_{cb}|$.} which allow us to derive the following bounds using the expressions from Ref.~\cite{Bobeth:2017ecx}, 
%%%%%%%%%%%%%%%
\begin{align}
\begin{split}
\dfrac{|y^{\prime\,2i}_{1L} \, y^{\prime\,1i\ast}_{1L}|}{M^2} &< (57~\mathrm{TeV})^{-2}\,,\qquad\quad i\in \lbrace 1,2,3\rbrace\,,
\end{split}
\end{align}
%%%%%%%%%%%%%%%     
     where $y^\prime \in \lbrace y^\prime_{1L},\,y^\prime_{2L},\,y^\prime_{3L}\rbrace$, and we have only considered the contributions that interfere with the SM ones. These constraints are weaker than the ones derived from $0\nu\beta\beta$ but still non-negligible for phenomenology as they allow us to probe all lepton flavors. Moreover, we do not consider constraints from $\Delta F=2$ processes on leptoquarks, as they are weaker due to the loop suppression, see e.g.~Ref.~\cite{Aebischer:2020dsw,Bobeth:2017ecx}. 
     
     \item[iii)] LFV $K$-decays: Leptoquarks also contribute at tree-level to LFV decays of kaons, which are subject to the very stringent constraints $\mathcal{B}(K_L\to \mu^\pm e^\mp)<4.7\times 10^{-12}$ and $\mathcal{B}(K^+\to \pi^+\mu^+ e^-)<1.3\times 10^{-11}$~(90$\%$ CL)~\cite{ParticleDataGroup:2022pth}. Using the EFT expressions from Ref.~\cite{Plakias:2023esq}, we derive the following upper bounds
%%%%%%%%%%%%%%%
\begin{align}
\begin{split}
   \dfrac{|y^{\prime\,21}_{2L} \, y^{\prime\,12\ast}_{2L}|}{M^2} &< (208~\mathrm{TeV})^{-2}\,,\qquad\qquad
   \dfrac{|y^{\prime\,21}_{3L} \, y^{\prime\,12\ast}_{3L}|}{M^2} < ({
   {290}}~\mathrm{TeV})^{-2}\,,
\end{split}
\end{align}
%%%%%%%%%%%%%%%        
     which are even more stringent than the above limits on lepton-flavor conserving kaon decays. Note, in particular, that the bounds on the $S_1$ leptoquark are not quoted above since they only contribute to these process at the one-loop level, cf.~Eq.~\eqref{eq:R2t-Yuk-bis}.   
 \end{itemize}

\noindent Notice that we did not consider observables with heavy quarks and the $\tau$-lepton, as they have a much weaker sensitivity than the ones obtained from lepton and kaon physics. Furthermore, limits on the Electric Dipole Moment (EDM) of the electron and neutron do not apply to our scenario since leptoquark couplings are assumed to be real and since right-handed couplings are not considered, cf.~Ref.\cite{Dekens:2018bci}. Finally, we have also estimated using naive dimensional analysis the leptoquark contributions to LNV processes such as $K^+\to \pi^- e^+ e^+$, and analogous decay modes, which are sensitive to the same combination of leptoquark couplings than $0\nu\beta\beta$. However, the current experimental limits turn out to be too weak to provide useful constraints on these scenarios~\cite{ParticleDataGroup:2022pth}, due to the $\lambda_{1(3)}/M^4$ suppression appearing in the $d=7$ SMEFT operators contributing to these processes, cf.~Eqs.~\eqref{eq:matching-R2-S1} and \eqref{eq:matching-R2-S3}.

%%%%%%%%%%%%%%%%%%%%%%%%%%%%%%%%%%%%%%%%
%%%%%%%%%%%%%%%%%%%%%%%%%%%%%%%%%%%%%%%%
\section{Numerical results}\label{sec:numerical}
%%%%%%%%%%%%%%%%%%%%%%%%%%%%%%%%%%%%%%%%
%%%%%%%%%%%%%%%%%%%%%%%%%%%%%%%%%%%%%%%%

In this Section, we study the correlation between neutrino masses and the $0\nu\beta\beta$ half-life in the two leptoquark models introduced above. We will assume that the leptoquark interactions are the only source of neutrino masses via the one-loop diagram depicted in Fig.~\ref{fig:diag-loop}. The trilinear couplings $\lambda_{1(3)}$ will be fixed to $M/2$, in agreement with the electroweak constraints from Sec.~\ref{sec:pheno}, where $M \equiv m_{S_{1(3)}} = m_{\widetilde{R}_2}$ is a free parameter. The Yukawas $y_{2L}$ and $y_L$ are assumed to be real and are varied within the perturbativity constraint ($\smash{\lesssim \sqrt{4\pi}}$), without imposing any specific flavor pattern other than the requirement that we must reproduce the $\Delta m_{ij}^2 \equiv m_{\nu_i}^2-m_{\nu_j}^2$ values determined experimentally within $3\sigma$ accuracy, cf.~Eq.~\eqref{eq:oscillation}. The viable Yukawas are then used to find the unitary matrix $U_{\nu_L}$ and to determine $U_{\ell_L}=U_{\nu_L} U^\dagger$. The PMNS matrix $U$ is parameterized as usual,
%%%%%%%%%%%%%%%%
\begin{align}
    U = \begin{pmatrix}
1 & 0 & 0\\ 
0 & c_{23} & s_{23}\\ 
0 & -s_{23} & c_{23} 
\end{pmatrix} 
\,
\begin{pmatrix}
c_{13} & 0 & s_{13} \, e^{-i\delta_{CP}}\\ 
0 & 1 & 0\\ 
s_{13} \, e^{i\delta_{CP}} & 0 &  c_{13}
\end{pmatrix}
\,
\begin{pmatrix}
c_{12} & s_{12} & 0\\ 
-s_{12} & c_{12} & 0\\ 
0 & 0 & 1
\end{pmatrix}
\,
\mathcal{P}\,,
\end{align}
%%%%%%%%%%%%%%%%
where $s_{ij}\equiv \sin\theta_{ij}$, $c_{ij}\equiv \cos\theta_{ij}$ and $\mathcal{P}=\mathrm{diag}(e^{i\alpha_1},e^{i\alpha_2},1)$
for a given pair of values ($\alpha_1,\alpha_2$) of the unknown Majorana phases, defined in the range $0 \leq \alpha_{1,2}\leq \pi$. We consider the results for the neutrino mass differences $\Delta m_{ij}^2$ from the latest update of Ref.~\cite{Esteban:2020cvm} for both Inverted Ordering  (IO) and Normal Ordering (NO) of neutrino masses,
%%%%%%%%%%%%%%%
\begin{align}
    \label{eq:oscillation}
    \Delta m_{21}^2 &= (7.4 \pm 0.2)\times 10^{-5}~\mathrm{eV}^2 \,, \qquad |\Delta m_{3\ell}^2| = (2.51 \pm 0.03)\times  10^{-3}~\mathrm{eV}^2\,,
\end{align}
%%%%%%%%%%%%%%%

\noindent with $\ell=1\, (2)$ for NO (IO) and we vary the absolute scale of neutrino masses, imposing the conservative upper limit $\sum_i m_{\nu_i} \lesssim 0.12$~eV from cosmology \cite{Planck:2018vyg}. For the PMNS matrix $U$, we consider the $3\sigma$ ranges provided in Ref.~\cite{Esteban:2020cvm},~\footnote{We use the NuFIT 2024 results, without Super-Kamiokande atmospheric data, which is provided in the web page \href{http://www.nu-fit.org/}{http://www.nu-fit.org/}.}
%%%%%%%%%%%%%%%
\begin{align}
|U|^{3\sigma}=\begin{pmatrix}
0.801 \to 0.842 & 0.518 \to 0.580 & 0.142 \to 0.155\\[0.35em] 
0.236 \to 0.507 & 0.458 \to 0.691 & 0.630 \to 0.779\\[0.35em]
0.264 \to 0.527 & 0.471 \to 0.700 & 0.610\to 0.762
\end{pmatrix}
\,.
\end{align}
%%%%%%%%%%%%%%%

\subsection{Inverted or normal ordering?}

%%%%%%%%%%%%%%
\begin{figure}[!t]
\begin{center}
\includegraphics[width=0.67\textwidth]{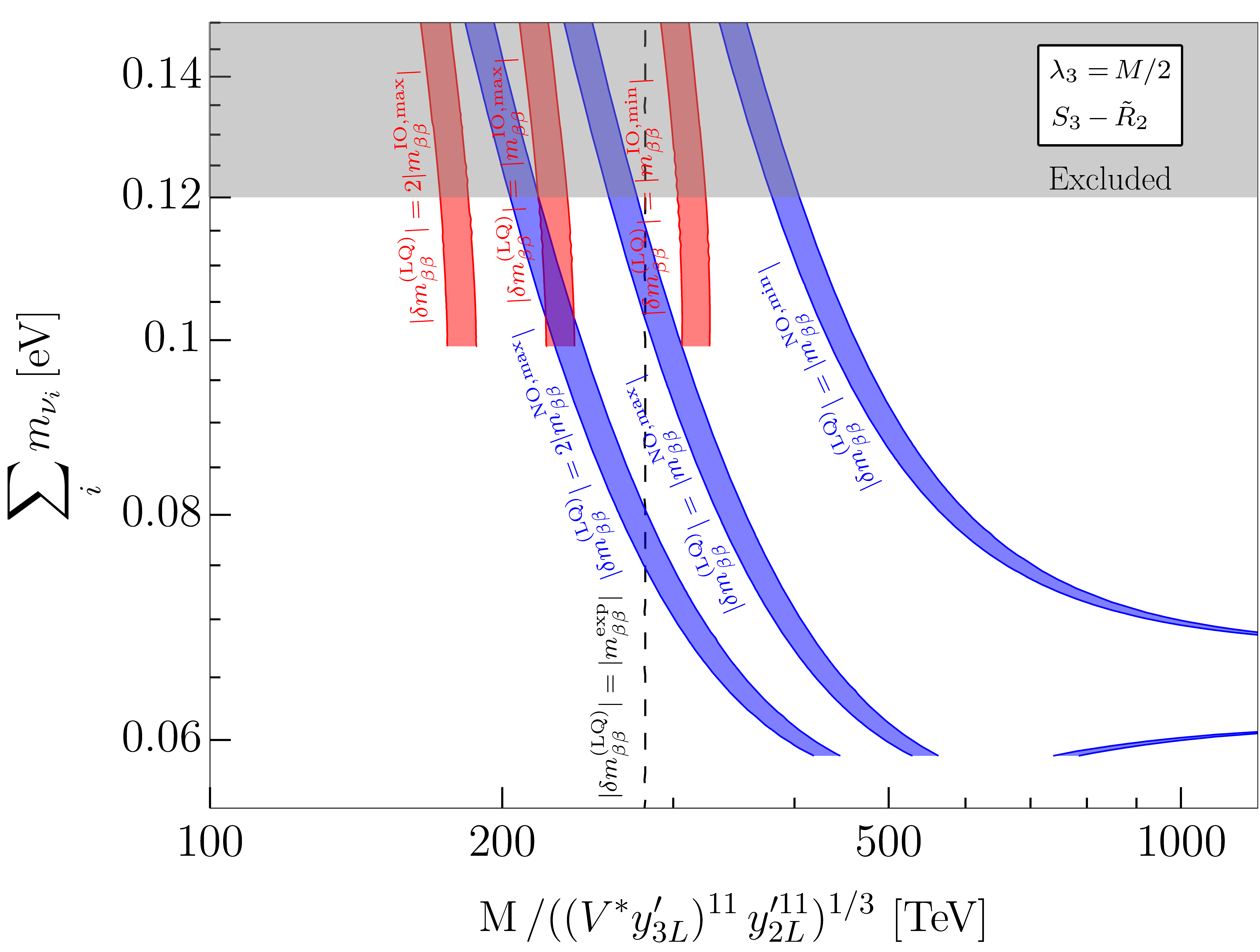} 
\end{center}
\caption{
\small \sl The contour lines for $\smash{\delta m_{\beta\beta}^{(\mathrm{LQ})}}$, defined in Eq.~\eqref{eq:mbb_eff}, are shown in the plane $\smash{\sum_i m_{\nu_i}}$ vs.~$\smash{M/(V^* y_{3L}^{\prime})^{11}\, y_{2L}^{\prime 11})^{1/3}}$ for the $\widetilde{R}_2$-$S_3$ scenario, for both NO (blue lines) and IO (red lines), by fixing $\lambda_3=M/2$. For illustration, we show the regions where $\smash{\delta m_{\beta\beta}^{(\mathrm{LQ})}} \in \lbrace m_{\beta\beta}^{\mathrm{min}},\,m_{\beta\beta}^{\mathrm{max}},\, 2m_{\beta\beta}^{\mathrm{max}}\rbrace$. We consider a degenerate mass $M$ for the leptoquark states and the primed leptoquark Yukawas, defined in Eq.~\eqref{eq:primed-yuk-S1}, to account for the leptonic mixing matrices. The thickness of the blue and red lines comes from the QCD RGE effects described in Sec.~\ref{sec:0nubb}.
\label{fig:plot-0vbb-LQ}
}
\end{figure}
%%%%%%%%%%%%%%

We now discuss our main numerical results for the correlation between neutrino masses and the effective Majorana $m_{\beta\beta}$ for both NO and IO. As we will demonstrate, the tree-level leptoquark contributions can interfere destructively or constructively with the contributions from Majorana masses, depending on the sign of the leptoquark couplings, which could decrease or increase the $0\nu\beta\beta$ decay rates. 

In Fig.~\ref{fig:plot-0vbb-LQ}, we first estimate the size of the leptoquark contributions to $m_{\beta\beta}^\mathrm{eff}$ in comparison to the standard $m_{\beta\beta}$, as a function of the generate leptoquark mass $M$. To this purpose, we show the $\smash{\delta m_{\beta\beta}^{(\mathrm{LQ})}}$ contour lines in the plane defined by $\smash{\sum_{i} m_{\nu_i}}$ and the mass scale $M$ normalized by the combination of couplings entering the $0\nu\beta\beta$ amplitude, cf.~Eq.~\eqref{eq:mbb_eff}.~\footnote{We consider the $\widetilde{R}_2$-$S_3$ scenario in Fig.~\ref{fig:plot-0vbb-LQ}, but the conclusions are very similar in the scenario with the $\widetilde{R}_2$-$S_1$ leptoquarks.} For illustration, we choose the contour lines corresponding to the maximal and minimal values of $m_{\beta\beta}$ for both NO and IO, as well as the scenario where the leptoquark contribution from leptoquark is twice as large as the maximal value.  From this plot, we see that scenarios with $|\delta m_{\beta\beta}^{(\mathrm{LQ})}| \gg m_{\beta\beta}$ are in principle possible for leptoquark masses as large as~$\approx 500$~TeV, for perturbative leptoquark couplings, in agreement with Fig.~\ref{fig:plot-0vbb-LQ}. 

\begin{figure}[!p]
\begin{center}
\includegraphics[width=0.49\textwidth]{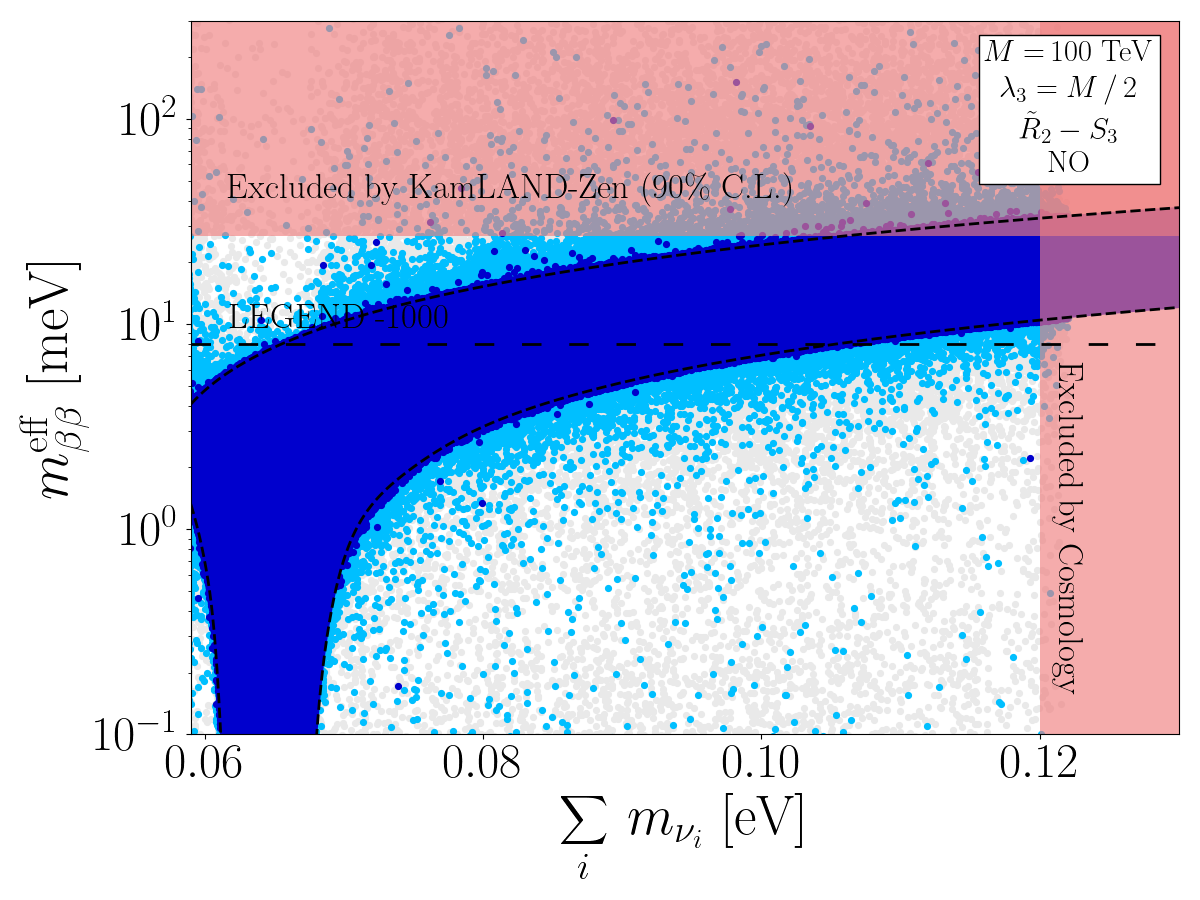}~\includegraphics[width=0.49\textwidth]{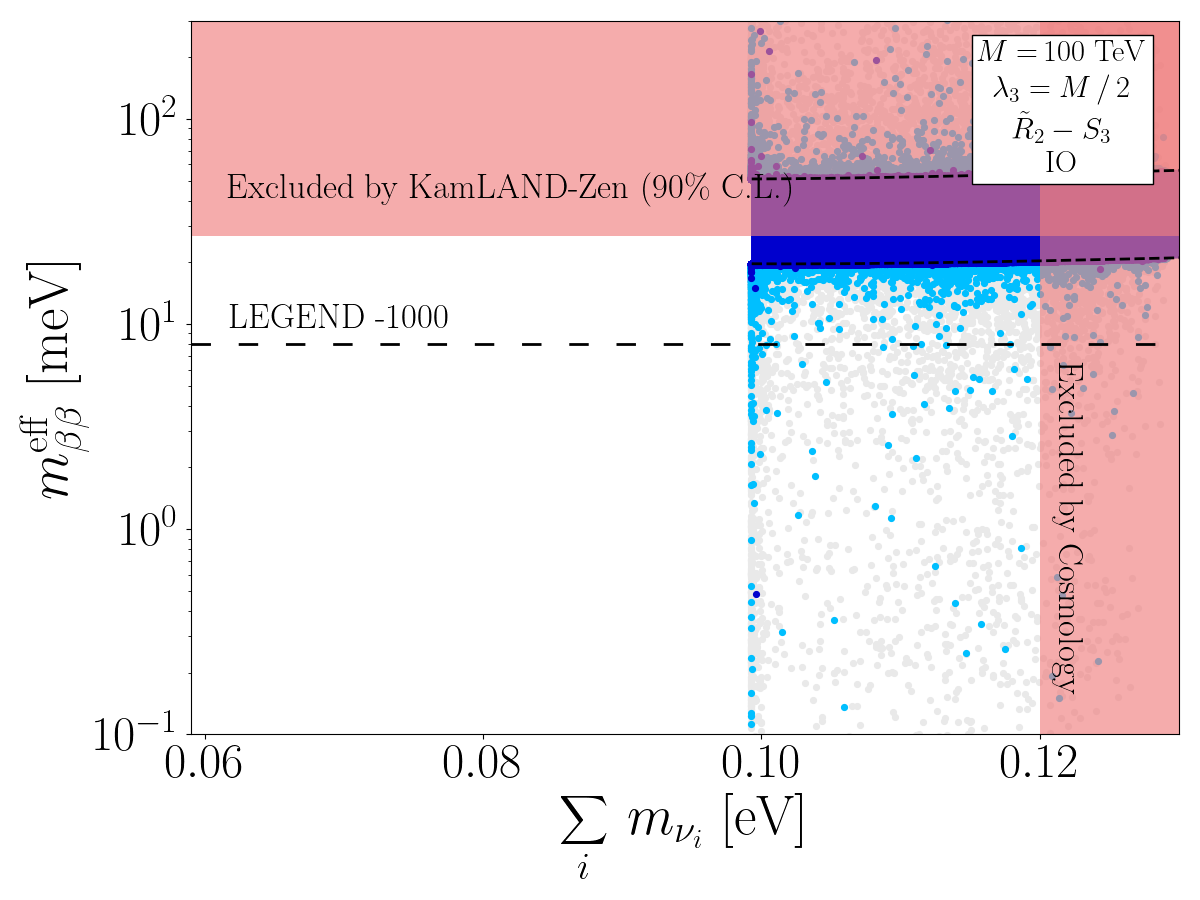}\\
\includegraphics[width=0.49\textwidth]{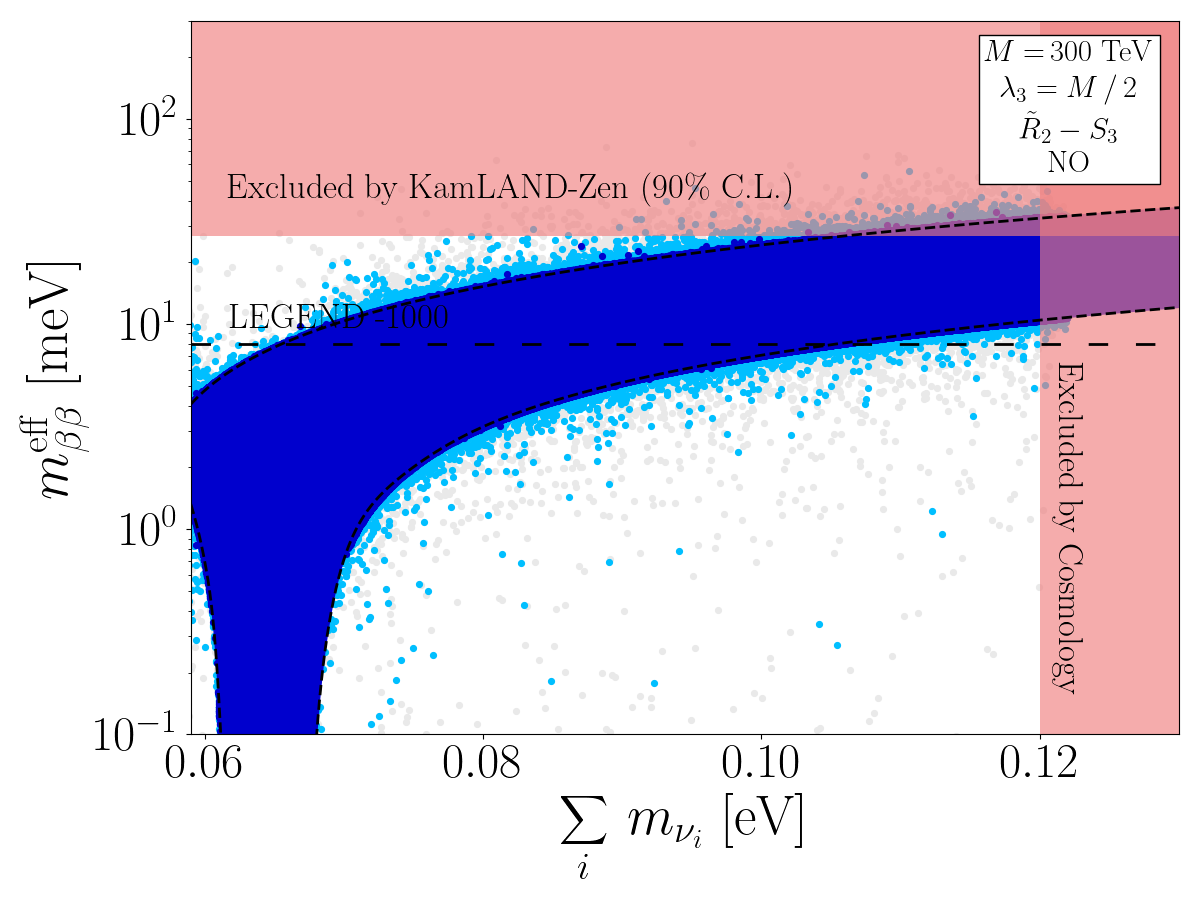}~\includegraphics[width=0.49\textwidth]{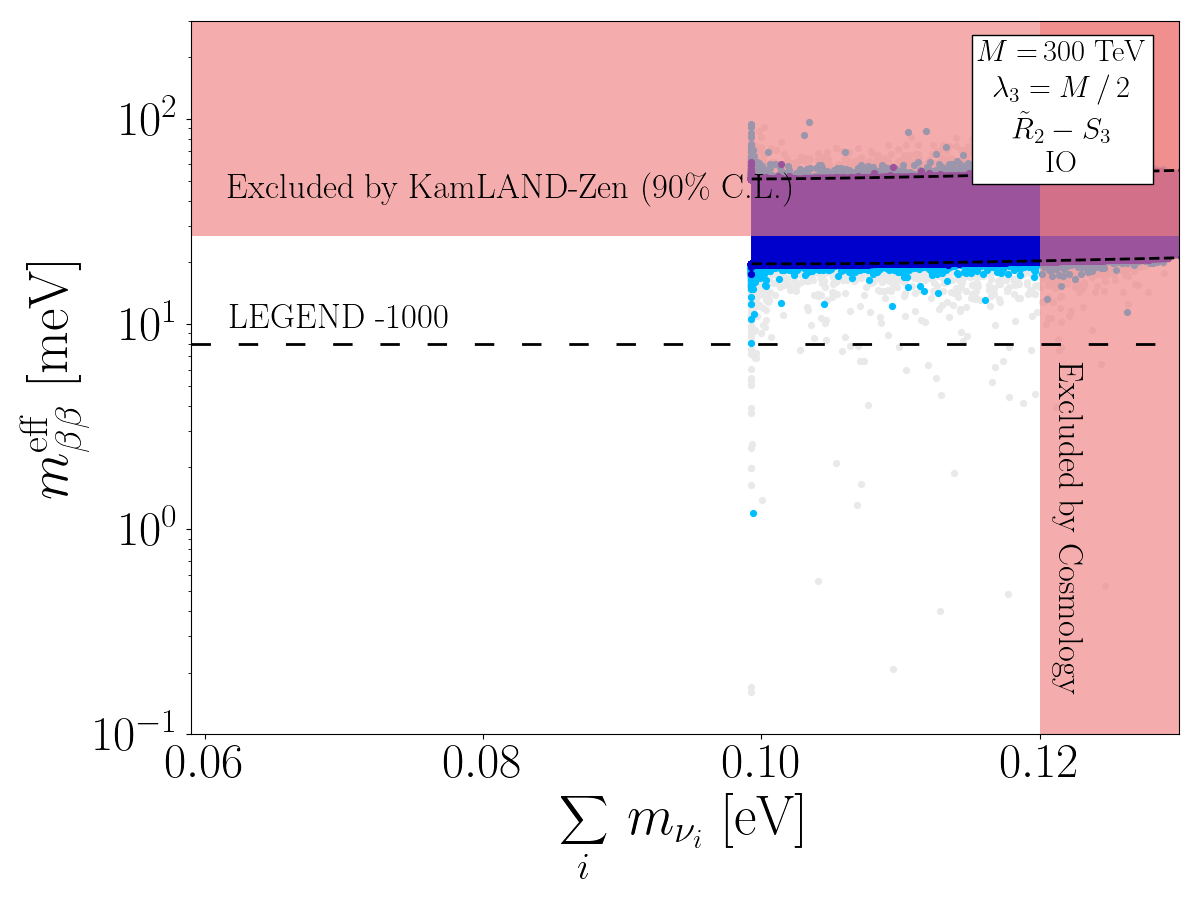}\\
\includegraphics[width=0.49\textwidth]{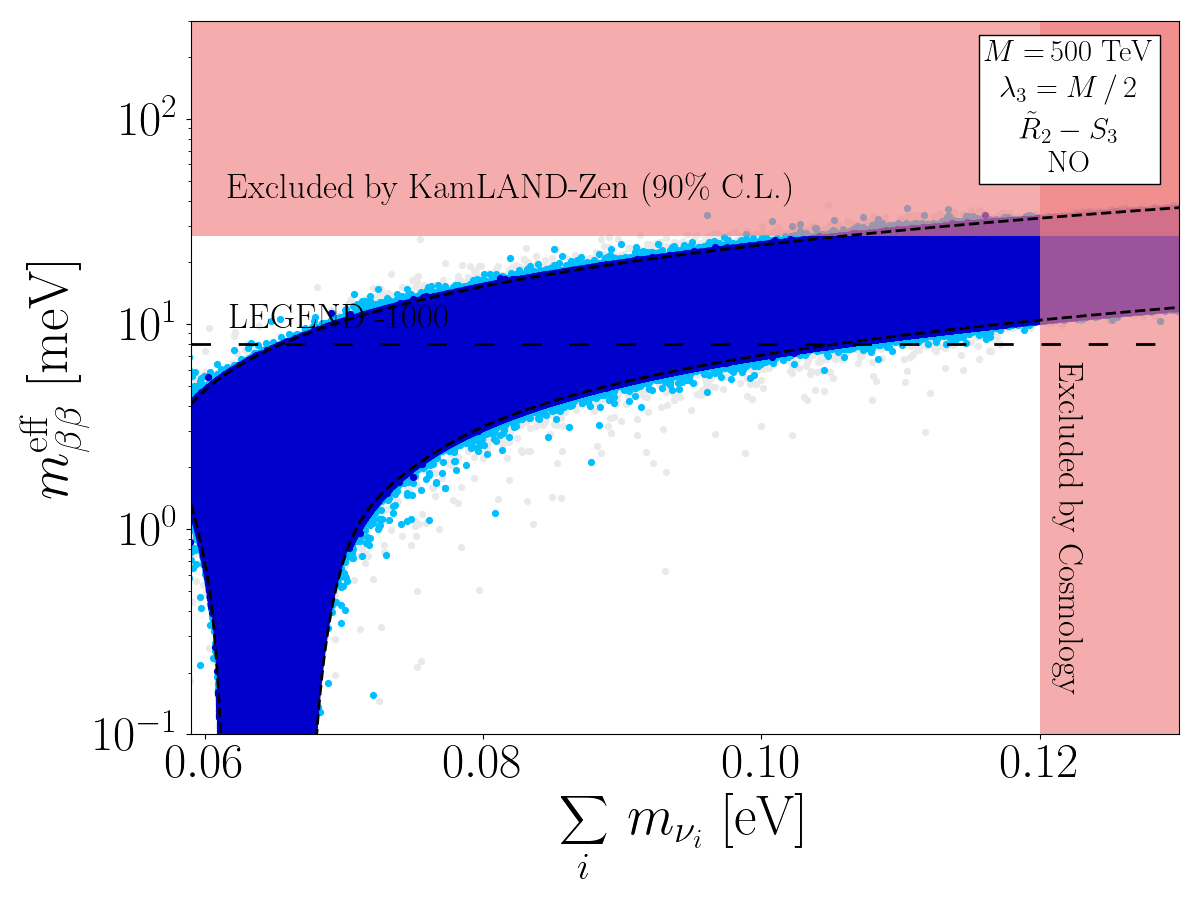}~\includegraphics[width=0.49\textwidth]{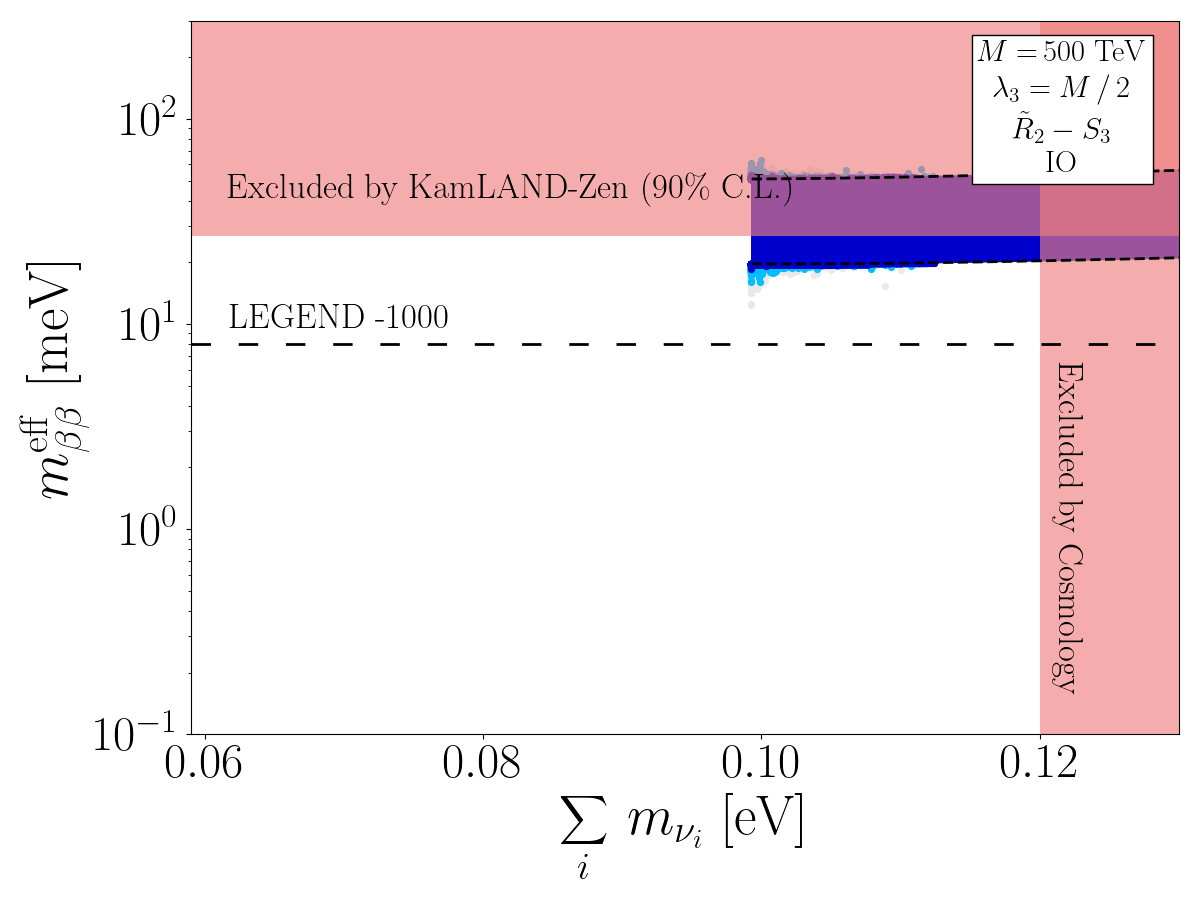}  
\end{center}
\caption{\small \sl Predictions for 
$m_{\beta\beta}^\mathrm{eff}$ defined in Eq.~\eqref{eq:mbb_eff} vs.~the sum of neutrino masses in the $S_3 - \widetilde{R}_2$ leptoquark model for the NO (left column) and IO (right column) scenarios. 
The points in gray (light-blue) are currently excluded by (in the reach of future data on) 
the flavor physics observables discussed in Sec.~\ref{sec:pheno}.
The leptoquark masses are fixed to be degenerate, with $M \in \lbrace 100, 300, 500\rbrace$ TeV, and the trilinear coupling is set to $\lambda_3 = M/2$, in agreement with the constraints discussed in Sec.~\ref{sec:0nubb}. The dashed lines delimit the region allowed if only the $d=5$ Weinberg operator is present. The current bounds from KamLAND-Zen~\cite{KamLAND-Zen:2024eml} and cosmology~\cite{Planck:2018vyg} are depicted in red.
We also show, as a reference, the expected reach of LEGEND-1000~\cite{LEGEND:2021bnm}.}
\label{fig:plot-S3}
\end{figure}

Based on the above discussion, we consider three benchmark values for the leptoquark masses in the $S_3$-$\widetilde{R}_2$ model, namely $M \in \lbrace 100,~300,~500\rbrace~\mathrm{TeV}$, with the other parameters varied as described above.~\footnote{Similar results can be obtained in the scenario with the singlet leptoquark $S_1$ instead of the triplet $S_3$.} In Fig.~\ref{fig:plot-S3}, we show the correlation between the $0\nu\beta\beta$ effective Majorana mass ($m_{\beta\beta}^\mathrm{eff}$) defined in Eq.~\eqref{eq:mbb_eff} and the sum of neutrino masses ($\sum_i {m_{\nu_i}}$) for the NO (left column) and IO (right column) scenarios. In this plot, the gray points are excluded by the flavor physics observables listed in Sec.~\ref{sec:pheno}, whereas the light blue ones are within reach of future experiments, as will be discussed below. The navy blue points are allowed by both current and projected bounds.
For masses smaller than $\approx 300$~TeV, we indeed find that $m_{\beta\beta}^{\mathrm{eff}}$ could be considerably decreased or increased with respect to $m_{\beta \beta}$ for either ordering, while remaining compatible with other constraints. For larger values of $M$, the additional contributions to $0\nu\beta\beta$ decrease, in agreement with the dimensional-analysis estimation from Eq.~\eqref{eq:nda}.

%%%%%%%%%%%%%%
%%%%%%%%%%%%%%

By comparing the normal and inverted ordering plots in Fig.~\ref{fig:plot-S3}, we also notice an important difference between the leptoquark models and the scenario where the only contribution to $0\nu\beta\beta$ arises from the Weinberg operator. In the latter case, there is a minimum value 
for $m_{\beta \beta}$  when $\alpha_1,\alpha_2$ are 
0 or $\pi/2$~\cite{Nunokawa:2002iv}. For the IO this implies the lower limit $m_{\beta\beta} \gtrsim \sqrt{\vert \Delta m^2_{32}\vert}
\cos 2\theta_{12} \sim 20~\mathrm{meV}$, whereas for the NO 
this minimum may even be zero as there can be an exact cancellation for a small range of values of $m_{\nu_1}$ consistent with $\sum_i  m_{\nu_i}\sim 0.06 $ eV. This important distinction between the two ordering scenarios is lifted once the leptoquark contributions to $m_{\beta\beta}^\mathrm{eff}$ are included since the standard contributions can interfere destructively with the ones induced by leptoquarks, cf.~Eq.~\eqref{eq:mbb_eff}. Therefore, in this case, it is fundamental to have independent information on the neutrino ordering from oscillation experiments. 

\paragraph{Resolving the ambiguity} We expect the medium baseline ($\sim$ 50 km) JUNO reactor neutrino experiment~\cite{JUNO:2022mxj}, by exploring $\bar\nu_e \to \bar \nu_e$ oscillations, to determine the neutrino mass ordering at  3$\sigma$ in about a decade. 
They can achieve this by a very precise measurement ($\sim$ 0.2\%) of $\vert \Delta m^2_{ee}\vert \equiv \Delta m^2_{31}\cos^2\theta_{12} + 
 \Delta m^2_{32}\sin^2\theta_{12}$, which will 
 differ at JUNO by $1.8 \times 10^{-5}$ eV$^2$ for NO and IO~\cite{Forero:2021lax}. 
 Recently, however, it has been shown that JUNO's result, when combined with the current $\nu_\mu \to \nu_\mu$ disappearance data from T2K and NOvA, can determine the neutrino mass ordering at 3$\sigma$ within one year of operation~\cite{Parke:2024xre}.
 Ultimately the future accelerator neutrino oscillation long baseline ($\sim$ 1300 km) experiment DUNE~\cite{DUNE:2020ypp}, by exploring matter effects in $\nu_ \mu \to \nu_e$ oscillations, should be able to determine the ordering at 5$\sigma$. The absolute scale of neutrino masses will probably only be determined in a laboratory experiment for the IO spectrum.  Neutrino oscillation experiments place lower limits on the effective neutrino mass $m_\beta \equiv \sqrt{\sum_i \vert U_{ei}\vert^2 m^2_{\nu_i}}$ that can be measured in $\beta$-decay experiments: $m_\beta \gtrsim 0.048$ eV (IO) and $m_\beta \gtrsim 0.0085$ eV (NO). 
 The current state-of-the-art $\beta$-decay experiment KATRIN is projected to probe  $m_\beta \gtrsim$ 0.2 eV~\cite{KATRIN:2022ayy}, while a new idea, the use of Cyclotron Radiation Emission Spectroscopy,
 promoted by the Project 8 experiment, may allow to 
 reach down to  $m_\beta \sim 0.04$ eV~\cite{Project8:2022wqh}. 
 If neutrinos have a NO spectrum, cosmology may be the only 
 realistic hope to access the absolute scale in the near 
 future  with CMB-S4\cite{Abazajian:2019eic}. 
  
\subsection{Implications to flavor observables}
\label{ssec:flavor-probes}

Finally, we discuss the implications of the viable scenarios considered above to flavor physics observables. As already discussed in Sec.~\ref{sec:pheno}, the most sensitive flavor bounds to the leptoquark couplings are kaon decays and $\mu N\to e N$ processes, which can be described by $d=6$ operators in the SMEFT~\cite{Buchmuller:1986zs}, cf.~Appendix~\ref{app:d6-matching}. These processes can probe masses as large as $\mathcal{O}(10^3~\mathrm{TeV})$ with current data, depending on the specific model, for perturbative leptoquark couplings. These bounds are translated in Fig.~\ref{fig:plot-S3} by the gray points that are excluded by flavor bounds, which are dominated by $\mu N \to e N$.

\begin{figure}[!t]
\begin{center}
\includegraphics[width=0.67\textwidth]{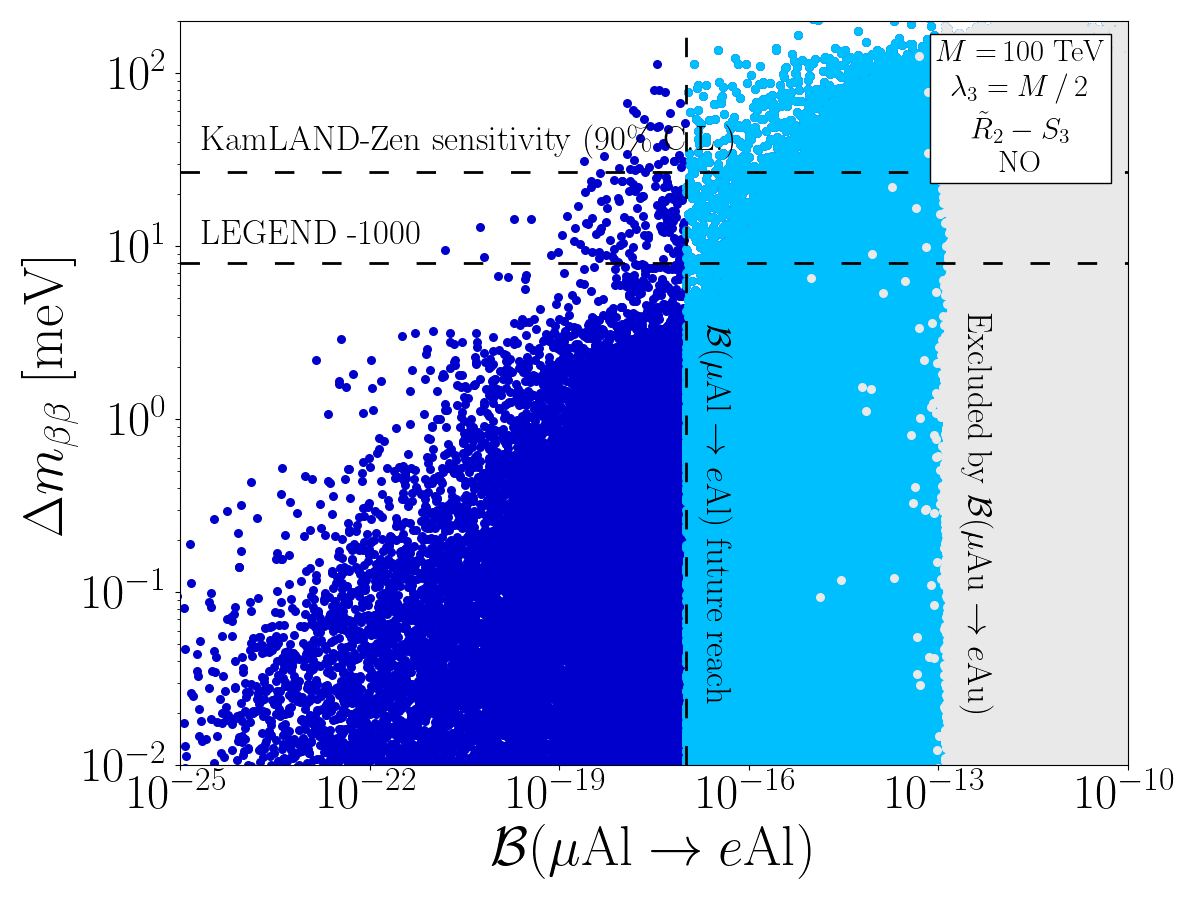} 
\end{center}
\caption{\small\sl
The leptoquark contributions to $\Delta m_{\beta\beta}\equiv m_{\beta\beta}^{\mathrm{eff}}-m_{\beta\beta}$ is plotted against $\mathcal{B}(\mu \mathrm{Al}\to e \mathrm{Al})$ for the $\smash{\widetilde{R}_2}$-$S_3$ model with degenerate masses $M=100$~TeV and the trilinear coupling fixed to $\lambda_3=M/2$, which corresponds to the top-left panel in Fig.~\ref{fig:plot-S3}. The gray points are excluded by the current best limit with gold atoms, $\mathcal{B}_{\mu e}^{(\mathrm{Au})}<7\times 10^{-13}$ (90~$\%$ CL)~\cite{ParticleDataGroup:2022pth}, the light-blue points are within reach of Mu2E~\cite{Mu2e:2014fns} and COMET~\cite{COMET:2018auw} experiments, which will have a $\mathcal{O}(10^{-17})$ sensitivity for aluminum atoms, and the navy-blue points are beyond the reach of these experiments. In the same plot, we show the current bound from KamLAND-Zen~\cite{KamLAND-Zen:2024eml} and expected sensitivity of LEGEND-1000~\cite{LEGEND:2021bnm} for $0\nu\beta\beta$ decays.}
\label{fig:flavor-corr}
\end{figure}

In the near future, the Mu2E experiment at Fermilab~\cite{Mu2e:2014fns} and COMET at J-PARC~\cite{COMET:2018auw} will remarkably improve the experimental sensitivity for $\mu \to e$ conversion in aluminum atoms, with an expected precision of $\mathcal{O}(10^{-17})$. These constraints would imply an improvement of about three orders of magnitude on the reach for leptoquark masses, cf.~Eq.~\eqref{eq:limits-muNeN}. The impact of these constraints is depicted by the light-blue points in Fig.~\ref{fig:plot-S3}, which can be fully probed by these experiments. We find that there are very few points which have a substantial leptoquark contribution that could lead to an observable signal in LEGEND-1000, while keeping the $\mu N\to eN$ rate below the sensitivity of future LFV experiments (i.e., navy-blue points in Fig.~\ref{fig:plot-S3} far from the region delimited by the dashed line). This feature is also depicted in Fig.~\ref{fig:flavor-corr} for the $\widetilde{R}_2$-$S_3$ model, where the $\mu N \to e N$ conversion rate is plotted against the difference $\Delta m_{\beta\beta} \equiv m_{\beta\beta}^\mathrm{eff}-m_{\beta\beta}$ that quantifies the leptoquark contributions to $0\nu\beta\beta$. In the following, we explore the reason for this correlation, which is connected to the requirement of explaining neutrino masses.

\paragraph*{$0\nu\beta\beta$ decays vs.~$\mu N\to eN$}
To understand the apparent correlation between the leptoquark contributions to $0\nu\beta\beta$ decays and $\mu N\to e N$ shown in Fig.~\ref{fig:flavor-corr}, we spell out the low-energy EFT describing the latter~\cite{Kitano:2002mt},
%%%%%%%%%%%%%
\begin{align}
\label{eq:left}
\mathcal{L}^{(\mu\rightarrow e)}_\mathrm{eff.} \supset \sqrt{2}\, G_F
 \sum_{q=u,d,s} \, \sum_{X,Y=L,R} C_{V_{XY}}^{(q)}\, \big{(}\bar{e} \gamma^\mu P_X \mu\big{)}\big{(} \bar{q} \gamma_\mu P_Y q\big{)}  + \mathrm{h.c.}\,,
\end{align}
%%%%%%%%%%%%%
where we neglected scalar operators and the ones with the gluon field strength, which are suppressed in our setup~\cite{Plakias:2023esq}. The nonzero coefficients $C_{V_{XY}}^{(q)}$ can be expressed at tree-level as
%%%%%%%%%%%%%
\begin{align}
\label{eq:smeft-matching-mu-e-1}
C_{V_{LL}}^{(u)} & = \dfrac{v^2}{2}\dfrac{(V^* y_{3L}^{\prime})_{12}(V^* y_{3L}^{\prime})_{11}^\ast}{m_{S_3}^2} \, + \, \dfrac{v^2}{2}\dfrac{(V^* y_{1L}^{\prime})_{12}(V^* y_{1L}^{\prime})_{11}^\ast}{m_{S_1}^2} \,,\\[0.3em]
\label{eq:smeft-matching-mu-e-2}
C_{V_{LL}}^{(d)} & = v^2 \dfrac{(y_{3L}^{\prime})_{12}(y_{3L}^{\prime})_{11}^\ast}{m_{S_3}^2} \, ,\\[0.3em]
\label{eq:smeft-matching-mu-e-3}
C_{V_{LR}}^{(d)} & = \,-\,\dfrac{v^2}{2}\dfrac{(y_{2L}^{\prime})_{12}(y_{2L}^{\prime})_{11}^\ast}{m_{\Tilde{R}_2}^2} \,.
\end{align}
%%%%%%%%%%%%%
These coefficients are then combined into the ones appearing in the nucleon EFT~\cite{Kitano:2002mt},
%%%%%%%%%%%%%
\begin{align}
c_{VL}^{(p)} &= \sum_{q} f_{Vp}^{(q)}  \,(C_{V_{LR}}^{(q)}+C_{V_{LL}}^{(q)})\,,\qquad\quad
c_{AL}^{(p)} = \sum_{q} f_{Vp}^{(q)}  \, (C_{V_{LR}}^{(q)}-C_{V_{LL}}^{(q)})\,,
\end{align}
%%%%%%%%%%%%%
\noindent and, similarly, for $p\to n$, where $\smash{f_{Vp}^{(u)}=f_{Vn}^{(d)}=2}$, $\smash{f_{Vp}^{(d)}=f_{Vn}^{(u)}=1}$ and $\smash{f_{Vp}^{(s)}=f_{Vn}^{(s)}=0}$ comes from the valence content of nucleons. The vector coefficients $c_{VL}^{(p)}$ and $c_{VL}^{(n)}$ are tightly constrained, as they give rise to the spin-independent rates~\cite{Davidson:2017nrp}, 
%%%%%%%%%%%%%%
\begin{equation}
    \label{eq:mueN-SI}
    \mathcal{B}_{\mu e}^{(N)}\Big{\vert}_{\mathrm{SI}} = \dfrac{2 \, G_F^2 \,m_\mu^5}{\Gamma_{\rm capt}^{N} \, \pi^2}\,(Z \alpha_{\rm em})^3 \left\vert Z \, F_p^{N}(m_\mu) \,c_{VL}^{(p)} + (A - Z) \, F_n^{N}(m_\mu) \,c_{VL}^{(n)}  \right \vert^2 + (L \leftrightarrow R)\,,
\end{equation}
%%%%%%%%%%%%%%
which are enhanced by the size of the nucleus $N$, where $Z$ denotes the atomic number and $A$ is the mass number. The capture rates are given by $\Gamma_{\rm capt}^{\mathrm{Al}} = 4.6 \times 10^{-16} \, {\rm MeV}$ and $\Gamma_{\rm capt}^{\mathrm{Au}} = 8.7 \times 10^{-15} \, {\rm MeV}$ for gold $(^{197}_{79} \mathrm{Au})$ and aluminum $(^{27}_{13} \mathrm{Al})$ atoms, respectively~\cite{Suzuki:1987jf}. The proton and neutron form factors are $F_p^{\rm Al}(m_\mu) = F_n^{\rm Al}(m_\mu) \approx 0.53$ and $F_p^{\rm Au}(m_\mu) = F_n^{\rm Au}(m_\mu) \approx 0.035$~\cite{Kitano:2002mt}. 

%%%%%%%%%%%%%%
%%%%%%%%%%%%%%

From the above equations, we can infer the condition for large $0\nu\beta\beta$ rates with suppressed $\mu N\to e N$ processes, as represented by the inaccessible region in the top-left corner of Fig.~\ref{fig:flavor-corr}.  Since the couplings to electrons directly enter $0\nu\beta\beta$, the absence of points in this region implies that (i) either the contributions from $S_{1(3)}$ and $\widetilde{R}_2$ can exactly cancel out in Eq.~\eqref{eq:mueN-SI}; (ii) or that the couplings to muons can vanish in the physical basis. While a cancellation between the different contributions Eq.~\eqref{eq:mueN-SI} is possible in both models, it typically implies large axial-vector coefficients $c_{AL}^{p(n)}$, which could still be probed via their spin-dependent contributions to $\mu N\to eN$~\cite{Davidson:2017nrp},
%%%%%%%%%%%%%%%%
\begin{equation}
    \label{eq:mueN-SD}
    \mathcal{B}_{\mu e}^{(N)}\Big{\vert}_{\mathrm{SD}} = \dfrac{8 \, G_F^2 \,m_\mu^5}{\Gamma_{\rm cap}^{N} \, \pi^2}\,(Z \alpha_{\rm em})^3 \dfrac{J_{N}+1}{J_{N}} \left\vert S_{N}^{(p)} \, c_{AL}^{(p)} + S_{N}^{(n)} \, c_{AL}^{(n)}  \right \vert^2 \dfrac{S_{N}(m_\mu)}{S_{N}(0)} + (L \leftrightarrow R)\, ,
\end{equation}
%%%%%%%%%%%%%%%%
\noindent which have been included in Figs.~\ref{fig:plot-S3} and \ref{fig:flavor-corr}. Here, $\alpha_{\rm em}$ is the fine-structure constant, $J_N$ is the nuclear angular momentum, the spin expectation values for aluminum at the zero-momentum transfer limit are denoted by $S_{\rm Al}^{(n)} \approx 0.030$ and $S_{\rm Al}^{(p)} \approx 0.34$~\cite{Klos:2013rwa}, and finite momentum corrections are taken into account through the ratio $S_{\rm Al} (m_\mu) / S_{\rm Al} (0) \approx 0.29 $ \cite{Engel:1995gw}. These contributions are certainly much smaller than the spin-independent ones, but they are sufficient to prevent a complete cancellation between the two leptoquark contributions to $\mu N\to e N$ as we do not observe in Fig.~\ref{fig:flavor-corr}. 

The remaining possibility is that the couplings to muons of both leptoquarks vanish in the physical basis for charged leptons, so that both vector and axial coefficients cancel out in the nucleon EFT. More precisely, this would imply $(y_{2L}^\prime)_{12}= (V^\ast y_{1L}^\prime)_{12} \approx 0$ for the model with $S_1$, and $(y_{2L}^\prime)_{12}= (V^\ast y_{3L}^\prime)_{12} = (y_{3L}^\prime)_{12} \approx 0$ for the model of $S_3$, where the leptoquarks couple to both down- and up-quarks. The main constraint to such scenarios is to reproduce viable neutrino masses and mixing. Indeed, by rotating the right-hand side of  Eq.~\eqref{eq:mnus} to the primed basis, it is possible to show that the left-handed size depends only on the PMNS matrix, the neutrino masses and the unknown Majorana phases. For the model with $S_1$, we are able to find a very fine-tuned solution that suppresses all couplings to muons, without spoiling neutrino parameters, e.g.,~by taking $(y_{1L}^\prime)_{22}=-(y_{1L}^\prime)_{12} V_{ud}/V_{us}$ and $(y_{1L}^\prime)_{32}=0$, which is however very rare and of little phenomenological interest. For the model with the $S_3$ leptoquark, we could not find such a solution due to the constraint from neutrino masses and mixing, which further confirms the findings of our numerical analysis in Fig.~\ref{fig:flavor-corr}.

%%%%%%%%%%%%%%%%%%%%%%%%%%%%%%%%%%%%%%%%
%%%%%%%%%%%%%%%%%%%%%%%%%%%%%%%%%%%%%%%%
\section{Conclusion}\label{sec:conclusion}
%%%%%%%%%%%%%%%%%%%%%%%%%%%%%%%%%%%%%%%%
%%%%%%%%%%%%%%%%%%%%%%%%%%%%%%%%%%%%%%%%

In this paper, we have performed a comprehensive analysis of neutrinoless double-beta decays  in scenarios where neutrino masses are dynamically generated via one-loop contributions from scalar leptoquarks. The simplest models are the ones with a pair of scalar leptoquarks, namely $\widetilde{R}_2=(\mathbf{3},\mathbf{2},1/6)$ combined with $S_1=(\bar{\mathbf{3}},\mathbf{1},1/3)$ or $S_3=(\bar{\mathbf{3}},\mathbf{3},1/3)$, which should break lepton number by two units via their mixing in the scalar sector~\cite{Chua:1999si}, with implications for both neutrino masses and $0\nu\beta\beta$ decays. The masses and mixing of these leptoquark states are fixed by the small neutrino masses, whereas their Yukawa couplings can be probed by various low-energy observables.

The most striking prediction of this setup are the additional tree-level contributions to $0\nu\beta\beta$. Using an Effective Field Theory approach~\cite{Cirigliano:2018yza}, we have quantified these effects, which can be described by $d=7$ operators invariant under the full SM gauge symmetry. These contributions can be sizable, as they are chirality-enhanced ($\propto E/m_\nu$)~\cite{Hirsch:1996qy}, with respect to the ones induced by Majorana neutrino masses ($m_\nu$), where $E\approx 100~\mathrm{MeV}$ is the nuclear energy scale. We have shown that they could have an observable effect for leptoquark masses below $\mathcal{O}(300~\mathrm{TeV})$, which could increase or decrease the $0\nu\beta\beta$ half-life, depending on the relative sign of the leptoquark couplings. This conclusion holds for both NO  and IO  of neutrino masses,  creating an ambiguity between the two scenarios if leptoquark contributions are present in addition to the $d=5$ Weinberg operator. This is particularly compelling since the future experimental efforts aim to extend the current half-life sensitivity to $0\nu\beta\beta$-decays by roughly two orders of magnitude, possibly leading to the first observation of these processes~\cite{LEGEND:2021bnm,nEXO:2021ujk}.

Clearly, the main difficulty of probing physics beyond the SM with $0\nu\beta\beta$ decays is the uncontrolled theoretical uncertainties of the nuclear matrix elements that can shift the prediction for the $0\nu\beta\beta$ half-life~\cite{Dolinski:2019nrj}, thus mimicking effects from operators beyond the $d=5$ Weinberg operator. While there is an ongoing theoretical effort for improving these nuclear determinations, e.g.~by using data from $2\nu\beta\beta$ transitions~\cite{Jokiniemi:2022ayc} and muon capture on nuclei~\cite{Jokiniemi:2024zdl}, it is important to provide additional insights through correlations with other observables. These quantities are, in certain cases, less affected by theoretical uncertainties, including specific flavor physics processes that we have explored.

Besides $0\nu\beta\beta$ decays, we have considered the most relevant theoretical and phenomenological constraints on the leptoquark couplings, coming from perturbative unitarity, electroweak precision data and flavor observables. In particular, we have shown that kaon decays and $\mu\to e$ conversion in nuclei impose strict limits on the leptoquark contributions, probing a similar range of leptoquark masses than $0\nu\beta\beta$ decays. 
For leptoquark couplings that successfully generate neutrino masses, we have shown in Fig.~\ref{fig:plot-S3} that current flavor constraints exclude a substantial part of the parameters that would induce large contributions to $0\nu\beta\beta$ decays, but still leave room for large contributions for specific combinations of leptoquark couplings. In the near future, the most important flavor processes will be $\mu\to e$ conversion in nuclei, which will be searched at Mu2E~\cite{Mu2e:2014fns} and COMET~\cite{COMET:2018auw}. We have shown that the projected experimental sensitivity will imply an improvement of about three orders of magnitude of the constraints on leptoquark masses. These processes are complementary to $0\nu\beta\beta$ decays, with the potential to exclude a sizable part of the parameter space which is within reach of e.g.~LEGEND-1000~\cite{LEGEND:2021bnm}.

Finally, we note that we have chosen leptoquark masses above $100~\mathrm{TeV}$, near to the maximum reach of $0\nu\beta\beta$ experiments, to restrict our flavor analysis since few observables can probe such large energy scales. Clearly, lower leptoquark masses can also generate viable neutrino masses for smaller leptoquark couplings, with potential signatures to a larger spectrum of flavor observables, as well as to the direct searches at the LHC, which are beyond the scope of this paper (see e.g.~Ref.~\cite{Graesser:2022nkv}).

%%%%%%%%%%%%%%%%%%%%%%%%%%%%%%%%%%%%%%%%
%%%%%%%%%%%%%%%%%%%%%%%%%%%%%%%%%%%%%%%%
\section*{Acknowledgment}\label{sec:acknoledgment}
%%%%%%%%%%%%%%%%%%%%%%%%%%%%%%%%%%%%%%%%
%%%%%%%%%%%%%%%%%%%%%%%%%%%%%%%%%%%%%%%%

We thank A.~Abada and S.~R.~Alcaraz for useful discussions. This project has received support from the European Union’s Horizon 2020 research and innovation programme under the Marie Skłodowska-Curie grant agreement N$^\circ$~860881-HIDDeN and N$^\circ$~101086085-ASYMMETRY and from the SPRINT/CNRS agreement supported by FAPESP under Contract No.~2023/00643-0. This work was also supported by ANR PIA funding: ANR-20-IDEES-0002. L.P.S.~Leal is fully financially supported by FAPESP under Contracts No. 2021/02283-6 and No. 2023/12330-7. S.~Fajfer acknowledges
financial support from the Slovenian Research Agency
(research core funding No.~P1-0035, J1-3013 and N1-
0321). R. Z. F. is partially supported by FAPESP under Contract No. 2019/04837-9, and by  Conselho Nacional de Desenvolvimento Científico e Tecnológico (CNPq). 
L.P.S.~Leal would like to thank the hospitality of 
the IJCLab Theory Group. 

\clearpage

%%%%%%%%%%%%%%%%%%%%%%%%%%%%%%%%%%%%%%%%
%%%%%%%%%%%%%%%%%%%%%%%%%%%%%%%%%%%%%%%%
%%%%%%%%%%%%%%%%%%%%%%%%%%%%%%%%%%%%%%%%
%%%%%%%%%%%%%%%%%%%%%%%%%%%%%%%%%%%%%%%%
%%%%%%%%%%%%%%%%%%%%%%%%%%%%%%%%%%%%%%%%
%%%%%%%%%%%%%%%%%%%%%%%%%%%%%%%%%%%%%%%%
\appendix

%%%%%%%%%%%%%%%%%%%%%%%%%%%%%%%%%%%%%%%%
%%%%%%%%%%%%%%%%%%%%%%%%%%%%%%%%%%%%%%%%
\section{SMEFT $d=6$ operators}
\label{app:d6-matching}
%%%%%%%%%%%%%%%%%%%%%%%%%%%%%%%%%%%%%%%%
%%%%%%%%%%%%%%%%%%%%%%%%%%%%%%%%%%%%%%%%

In this Appendix, we provide the tree-level matching of the leptoquark models to the $d=6$ SMEFT Lagrangian that is used to derive the flavor bounds in Sec.~\ref{sec:pheno},
%%%%%%%%%%%%%%%%
\begin{align}
%%%%%%
\dfrac{1}{\Lambda^2}\mathcal{C}_{\substack{lq\\ijkl}}^{(1)} &= \dfrac{3 y_{3L}^{lj}\,y_{3L}^{ki\ast}}{4m_{S_3}^2} + \dfrac{y_{1L}^{lj}\,y_{1L}^{ki\ast}}{4m_{S_1}^2}\,, &
\dfrac{1}{\Lambda^2}\mathcal{C}_{\substack{lequ\\ijkl}}^{(3)} &= -\dfrac{y_{1R}^{lj}\,y_{1L}^{ki\ast}}{8m_{S_1}^2}\,,
\\[0.35em]
%%%%%%
\dfrac{1}{\Lambda^2}\mathcal{C}_{\substack{lq\\ijkl}}^{(3)} &= \dfrac{y_{3L}^{lj}\,y_{3L}^{ki\ast}}{4m_{S_3}^2}-\dfrac{y_{1L}^{lj}\,y_{1L}^{ki\ast}}{4m_{S_1}^2}\,,&
\dfrac{1}{\Lambda^2}\mathcal{C}_{\substack{lequ\\ijkl}}^{(1)} &=\dfrac{y_{1R}^{lj}\,y_{1L}^{ki\ast}}{2m_{S_1}^2} \,, \\[0.35em]
%%%%%%
\dfrac{1}{\Lambda^2}\mathcal{C}_{\substack{eu\\ijkl}} &= \dfrac{ y_{1R}^{lj}\,y_{1R}^{ki\ast}}{2m_{S_1}^2} \,, &
\dfrac{1}{\Lambda^2}\mathcal{C}_{\substack{ld\\ijkl}} &= -\dfrac{ y_{2L}^{kj}\,y_{2L}^{li\ast}}{2m_{\widetilde{R}_2}^2} \,,
\end{align}
%%%%%%%%%%%%%%%%

\noindent where we use the same notation of Ref.~\cite{Jenkins:2013zja}.~\footnote{The one-loop matching of $S_1$ and $S_3$ to the SMEFT can be found in  Ref.~\cite{Gherardi:2020det}).}

%%%%%%%%%%%%%%%%%%%%%%%%%%%%%%%%%%%%%%%%
%%%%%%%%%%%%%%%%%%%%%%%%%%%%%%%%%%%%%%%%
\section{Nuclear-physics inputs for $0\nu\beta\beta$}
\label{app:nuclear-inputs}
%%%%%%%%%%%%%%%%%%%%%%%%%%%%%%%%%%%%%%%%
%%%%%%%%%%%%%%%%%%%%%%%%%%%%%%%%%%%%%%%%

%%%%%%%%%%%%%%%%%
\begin{table}[!t]
\renewcommand{\arraystretch}{1.3}
\centering
\begin{tabular}{|cc||cc||cc||cc|}
\hline
\multicolumn{2}{|c||}{$n \to p e \nu \,,\, \pi \to e \nu$}  & \multicolumn{2}{c||}{$\pi\pi \to e e $}    & \multicolumn{2}{c||}{$n \to p \pi e e$}    & \multicolumn{2}{c|}{$n n \to ppee$}    \\\hline\hline
\multicolumn{1}{|c|}{$g_A$} & 1.271 $\pm$ 0.002 & \multicolumn{1}{c|}{$g_T^{\pi\pi}$} & 0  & \multicolumn{1}{c|}{$g_{VL}^{\pi N}$} & 0 & \multicolumn{1}{c|}{$g_{VL}^{N N}$} & 0 \\
\multicolumn{1}{|c|}{$g_M$} & 4.7 & \multicolumn{1}{c|}{} &  & \multicolumn{1}{c|}{$g_{T}^{\pi N}$} & 0 & \multicolumn{1}{c|}{$g_{T}^{N N}$} & 0 \\
\multicolumn{1}{|c|}{$g_T$} & 0.99  $\pm$ 0.06 & \multicolumn{1}{c|}{} &  & \multicolumn{1}{c|}{} &  & \multicolumn{1}{c|}{$g_\nu^{N N}$} & ${\rm - (92.9 \pm 46.5)\, GeV}^{-2}$ \\
\multicolumn{1}{|c|}{$g_T^\prime$} & 0 & \multicolumn{1}{c|}{} &  & \multicolumn{1}{c|}{} &  & \multicolumn{1}{c|}{$g_{VL,VR}^{E,m_e}$} & 0 \\
\multicolumn{1}{|c|}{$B$} & {\rm $\approx$ 2.7 GeV} & \multicolumn{1}{c|}{} &  & \multicolumn{1}{c|}{} &  & \multicolumn{1}{c|}{} &  \\ \hline
\end{tabular}
\caption{\label{tab:LECs}\small \sl Low-energy constants used to evaluate the $0\nu\beta\beta$ half-life \cite{Cirigliano:2018yza,Scholer:2023bnn}.}
\end{table}
%%%%%%%%%%%%%%%%%

For the reader's convenience, we collect in this appendix the definition of the NMEs as well as the numerical inputs used in our analysis of $0\nu\beta\beta$ in Sec.~\ref{sec:0nubb}. The low-energy constants that we used are listed in Table~\ref{tab:LECs}, where the ones set to zero correspond to parameters that have not yet been determined theoretically, as chosen in the default setting of the package provided in Ref.~\cite{Scholer:2023bnn}. In Tables~\ref{tab:NMEs} and \ref{tab:PSFs}, we list the nuclear matrix elements and the phase-space factors that we considered, taken from Ref.~\cite{Menendez:2017fdf} and \cite{Horoi:2017gmj}, respectively. These parameters are then used to compute the NMEs via~\cite{Cirigliano:2018yza}
%%%%%%%%%%%%%
\begin{align}
    M_{GT} &=  M_{GT}^{AA} + M_{GT}^{AP} + M_{GT}^{PP} +  M_{GT}^{MM}\, ,\\[0.3em]
    M_{T} \; \; &=  M_{T}^{AP} + M_{T}^{PP} + M_{T}^{MM} \, ,\\[0.3em]
    M_{PS}  &=  \frac{1}{2} M_{GT}^{AP} + M_{GT}^{PP} + \frac{1}{2} M_{T}^{AP} + M_{T}^{PP}\, ,\\[0.3em]  
    M_{T6}  &=  2\frac{g_T^\prime-g_T^{NN}}{g_A^2} \frac{m_\pi^2}{m_N^2} M_{F,sd} - 8 \frac{g_T}{g_M} (M_{GT}^{MM}+ M_T^{MM}) + g_T^{\pi N} \frac{m_\pi^2}{4m_N^2}(M_{GT,sd}^{AP}+ M_{T,sd}^{AP}) \nonumber \\ 
     &+  g_T^{\pi \pi} \frac{m_\pi^2}{4m_N^2}(M_{GT,sd}^{PP}+ M_{T,sd}^{PP}) \, , \\[0.3em]   
     M_{M} & = 2 \frac{g_A}{g_M}(M_{GT}^{MM}+M_T^{MM}) + \frac{m_\pi^2}{m_N^2}\left( - \frac{2}{g_A^2} g^{NN}_{VL} M_{F,sd} + \frac{1}{2} g^{\pi N}_{VL} (M_{GT,sd}^{AP}+M_{T,sd}^{AP})\right) \, , \\[0.3em]
     M_{m_e,L} & = \frac{1}{6} \left( \frac{g_V^2}{g_A^2} M_{F} - \frac{1}{3} \, (M_{GT}^{AA}- 4 M_{T}^{AA}) -3 \, (M_{GT}^{AP}+ M_{GT}^{PP} + M_T^{AP}+M_T^{PP})\right . \nonumber \\
     &\left . - 12 \, \frac{g_{VL}^{m_e}}{g_A^2} M_{F,sd}\right) \, , \\[0.3em]
      M_{m_e,R} & = \frac{1}{6} \left( \frac{g_V^2}{g_A^2} M_{F} + \frac{1}{3} \, (M_{GT}^{AA}- 4 M_{T}^{AA}) +3 \, (M_{GT}^{AP}+ M_{GT}^{PP} + M_T^{AP}+M_T^{PP})\right . \nonumber \\
     &\left . - 12 \, \frac{g_{VR}^{m_e}}{g_A^2} M_{F,sd}\right) \, , \\[0.3em]
     M_{E,L} & = -\frac{1}{3} \left( \frac{g_V^2}{g_A^2} M_F + \frac{1}{3} (2 M_{GT}^{AA}+M_T^{AA}) + 6 \frac{g_{VL}^E}{g^2_A} M_{F,sd}\right) \, , \\[0.3em]
      M_{E,R} & = -\frac{1}{3} \left( \frac{g_V^2}{g_A^2} M_F - \frac{1}{3} (2 M_{GT}^{AA}+M_T^{AA}) + 6 \frac{g_{VR}^E}{g^2_A} M_{F,sd}\right) \, .
\end{align}
%%%%%%%%%%%%%

%%%%%%%%%%%%%%%%%
\begin{table}[!t]
\renewcommand{\arraystretch}{1.3}
\centering
\begin{tabular}{|c|c|c|c|c|c|c|c|}
\hline
$M_F$ & $M_{GT}^{AA}$ & $M_{GT}^{AP}$ & $M_{GT}^{PP}$ & $M_{GT}^{MM}$ & $M_{T}^{AA}$ & $M_{T}^{AP}$ & $M_{T}^{PP}$ \\ \hline
-0.54 & 2.45 & -0.79 & 0.25 & 0.19 & 0.00 & 0.01 & -0.01  \\ \hline\hline
$M_{T}^{MM}$ & $M_{F,sd}$ & $M_{GT,sd}^{AA}$ & $M_{GT,sd}^{PP}$ & $M_{T,sd}^{AP}$ & $M_{T,sd}^{AP}$ & $M_{T,sd}^{PP}$ & \\ \hline
0.00 & -1.28 & 4.21 & -1.97 & 0.74 & 0.05 & -0.02 &   \\ \hline
\end{tabular}
\caption{\label{tab:NMEs} \small \sl Nuclear matrix elements of $^{136}{\rm Xe}$ obtained using shell model \cite{Menendez:2017fdf}.}
\end{table}
%%%%%%%%%%%%%%%%%

%%%%%%%%%%%%%%%%%
\begin{table}[!t]
\renewcommand{\arraystretch}{1.3}
\centering
\begin{tabu}{|c|c|c|c|c|c|}
\hline
$G_{01}$ & $G_{02}$ & $G_{03}$ & $G_{04}$ & $G_{06}$ & $G_{09}$ \\ \hline\hline
2.09 & 5.15 & 1.40 & 1.88 & 2.86 & 4.59 \\ \hline
\end{tabu}
\caption{\label{tab:PSFs}\small\sl Phase space factors of $^{136}{\rm Xe}$ given in units of $10^{-14} {\, \rm yr}^{-1}$~\cite{Horoi:2017gmj}. }
\end{table}
%%%%%%%%%%%%%%%%%

%%%%%%%%%%%%%%%%%%%%%%%%%%%%%%%%%%%%%%%%%%%%%%%%%%%%%%
%%%%%%%%%%%%%%%%%%%%%%%%%%%%%%%%%%%%%%%%%%%%%%%%%%%%%%

%%%%%%%%%%%%%%%%%%%%%%%%%%%%%%%%%%%%%%%%%%%%%%%%%%%%%%
%%%%%%%%%%%%%%%%%%%%%%%%%%%%%%%%%%%%%%%%%%%%%%%%%%%%%%

\end{document}